\documentclass[twocolumn]{preprint}

\begin{document}
\twocolumn[{%
\preprinttitle
{Nanoparticles with a BANG: Detonation-Driven Synthesis of Metals and Metal Oxides}%
{\preprintauthor{1,2}{Jan Maurycy Uszko}{https://orcid.org/0009-0003-4684-185X},
\preprintauthor{3}{Robert Matyáš}{https://orcid.org/0000-0002-8049-0315},
\preprintauthor{2}{Naoto Shirahata}{https://orcid.org/0000-0002-1217-7589},
\preprintauthor{4}{Stephen J. Eichhorn}{https://orcid.org/0000-0003-4101-273X},
\preprintauthor{5}{Avinash J. Patil}{https://orcid.org/0000-0002-1959-5310},
\preprintauthor{5}{Simon R. Hall}{https://orcid.org/0000-0002-2816-0191}}%
{
\affiliation{1}{Bristol Composites Institute (BCI), School of Civil, Aerospace and Design Engineering, University of Bristol, University Walk, Bristol, BS8 1TR, United Kingdom}
\affiliation{2}{Research Center for Materials Nanoarchitectonics (MANA), National Institute for Materials Science (NIMS), 1-2-1 Sengen, Tsukuba, Ibaraki, 305-0047, Japan}
\affiliation{3}{Institute of Energetic Materials, Faculty of Chemical Technology, University of Pardubice, Pardubice, Czech Republic}
\affiliation{4}{Henry Royce Institute for Advanced Materials, School of Materials, The University of Manchester, Oxford Rd, Manchester, M13 9PL, United Kingdom}
\affiliation{5}{School of Chemistry, University of Bristol, Cantock's Close, Bristol, BS8 1TS, United Kingdom}
}
\frontnote{Correspondence}{Simon Hall (\href{mailto:simon.hall@bristol.ac.uk}{simon.hall@bristol.ac.uk})}
\preprintabstract{Bristol Accelerated Nanoparticle Generation (BANG) method is a solvent- and capping-agent-free nanoparticle synthesis. Previous work was limited mostly to gold, silver and copper particles.  Here, the method is extended to ten further elements, including Pb, Bi, Tl, Mn, Fe, Zn, Pt, Ni, Cd and Pd. Detonation in air favoured metal formation for less oxophilic metals, whereas detonation of silver azide under nitrogen suppressed detectable silver oxide formation while retaining locally identified hexagonal close-packed (HCP) Ag alongside face-centred cubic (FCC) Ag. The outcome of the synthesis can be modified by changing the detonation environment, while the highly non-equilibrium conditions provide access to phases that are difficult to obtain using conventional synthetic routes.

BANG is a broadly applicable, highly non-equilibrium nanoparticle synthesis method whose products are governed by precursor chemistry and atmosphere, with potential access to kinetically trapped phases and direct supported-particle deposition.}
\frontnote{Keywords}{Nanoparticles \textbar{} Energetic materials \textbar{} Nonequilibrium processes \textbar{} Electron microscopy \textbar{} Allotropy}
\frontnote{Funding}{This work was funded by the EPSRC Centre for Doctoral Training in Composites Science, Engineering and Manufacturing (EP/S021728/1). A part of this work was supported by JSPS KAKENHI Grant Number 24K01462 and the Advanced Research Infrastructure for Materials and Nanotechnology in Japan (ARIM) of the Ministry of Education, Culture, Sports, Science and Technology (MEXT), Proposal Number JPMXP1225NM5200.}
\frontnote{Abbreviations}{BANG, Bristol Accelerated Nanoparticle Generation;
DI, deionised;
DTA, differential thermal analysis;
EDX, energy-dispersive X-ray spectroscopy;
EEW, electrical explosion of wires;
FCC, face-centred cubic;
FFT, fast Fourier transform;
FeCP, iron tris(carbohydrazide) perchlorate;
FT-IR, Fourier-transform infrared spectroscopy;
HCP, hexagonal close-packed;
HRTEM, high-resolution transmission electron microscopy;
ICSD, Inorganic Crystal Structure Database;
IR, infrared;
MnCP, manganese tris(carbohydrazide) perchlorate;
NHN, nickel tris(hydrazine) dinitrate;
NiCP, nickel tris(carbohydrazide) perchlorate;
NiEP, nickel tris(ethylenediamine) perchlorate;
PSD, position-sensitive detector;
PXRD, powder X-ray diffraction;
SEM, scanning electron microscopy;
TACP, tetraamminecopper perchlorate;
TEM, transmission electron microscopy;
ZnAzTZ, zinc 5,5'-azotetrazolate dihydrate.}
\medskip
}]

\section{Introduction}
Rapid energy release provides a route to nanoparticle formation under conditions far from equilibrium. Such processes can couple precursor conversion, nucleation and particle growth within a short reaction period, giving the potential to obtain crystalline products without solvents or added capping agents during the particle-forming step. Establishing their synthetic scope requires understanding of which precursor chemistries produce particles, what compositions and structures are produced and whether particle formation can be combined with deposition onto a support.

Metal-containing energetic compounds are distinctive precursors because they can supply both the metal source and the chemical energy for particle formation. Detonation of fulminating gold was recently shown to produce crystalline gold nanoparticles, laying the groundwork for an approach subsequently termed Bristol Accelerated Nanoparticle Generation (BANG).~\cite{Uszko2024Gold,Uszko2025Silver} Related studies demonstrated particle formation from copper and silver azides, while subsequent work extended the approach to other silver-containing energetic compounds.~\cite{Cen_2024,Uszko2025Silver} These studies established the feasibility of the method but examined a limited range of metals. Whether BANG can accommodate more diverse metal chemistries and how that diversity is reflected in the recovered products remain central questions.

A useful comparison is the electrical explosion of wires (EEW), in which an electrical discharge rapidly heats a metal source, followed by cooling and particle formation.~\cite{Kotov_2003,Karioris_1962} In BANG, the energy is instead supplied by decomposition of the energetic precursor. Both approaches can produce broad particle-size distributions and chemically heterogeneous products, making characterisation of the recovered material essential. For BANG, oxygen-bearing precursor constituents and the surrounding atmosphere may both contribute to product chemistry. Consequently, the formation of metal or metal oxide particles must be established experimentally rather than inferred from the identity of the metal alone. Detonation-derived silver also raises questions about the structures accessible through this approach. Earlier work identified predominantly FCC Ag together with silver oxide and local structures assigned to HCP Ag.~\cite{Uszko2025Silver} These observations motivate further examination of the relationship between synthesis conditions and the recovered structures.

Here, we extend BANG to precursors containing ten additional metals: Pb, Bi, Tl, Mn, Fe, Zn, Pt, Ni, Cd and Pd. Electron microscopy, elemental analysis and local structural measurements reveal crystalline metal-containing particles across this precursor library, with evidence supporting elemental metal and metal oxide assignments. We compare products from three nickel precursors, examine silver-containing particles obtained under air and nitrogen conditions, and demonstrate direct deposition of Pd-containing particles onto activated carbon, for potential use in catalytic processes. Together, these results establish the broader chemical scope of BANG and its capacity to combine particle formation with support deposition.

\section{Results \& Discussion}
\subsection{Generality of detonation-driven nanoparticle formation}
~\autoref{tab:bigtab} shows an overview of the results presented herein and in existing literature, specifying the main metal element, explosive precursors, as well as the proposed composition of particles produced. Particles created are highly crystalline, heterogeneous in size and polydisperse. This is consistent with earlier reports of particles produced using the BANG method and with their rapid formation. The elemental and structural observations support incorporation of precursor metals into the collected particles either as a metal or a metal oxide.

\begin{table*}[t]
    \centering
    \caption{Overview of nanoparticles prepared with the BANG method}
    \label{tab:bigtab}
\begin{tabular}{ccc}
Element             & Precursor                             & Identified Phases                         \\ \hline\hline
\groupedcell{0.5}{Pb} & Lead azide                            & \groupedcell{0.5}{Lead oxides}                   \\
                    & Lead picrate                          &                                                \\
Bi                  & Basic bismuth 5,5'-azotetrazolate           & Bi$_2$O$_3$                                    \\
Cd                  & Cadmium 5,5'-azotetrazolate complex           & Cd                                             \\
Tl                  & Thallium azide                        & Thallium oxides                         \\ \hline
Mn                  & Manganese tris(carbohydrazide) perchlorate  & Mn$_3$O$_4$                                    \\
\groupedcell{1.0}{Ni} & Nickel tris(hydrazine) dinitrate         & \groupedcell{1.0}{NiO}                           \\
                    & Nickel tris(ethylenediamine) perchlorate &                                                \\
                    & Nickel tris(carbohydrazide) perchlorate     &                                                \\
Fe                  & Iron tris(carbohydrazide) perchlorate       &  Iron oxides                              \\
\groupedcell{0.5}{Cu} & Tetraamminecopper perchlorate         & \groupedcell{0.5}{Copper oxides}                           \\
                    & Cuprous acetylide                     &                                                \\
Zn                  & Zinc 5,5'-azotetrazolate dihydrate              & Zinc oxides                                        \\ \hline
\groupedcell{2.0}{Ag} & Silver azide*                         & FCC Ag, HCP Ag                                 \\
                    & Silver azide~\cite{Uszko2025Silver}        & \groupedcell{1.5}{FCC Ag, HCP Ag, silver oxides} \\
                    & Silver fulminate~\cite{Uszko2025Silver}     &                                                \\
                    & Silver acetylide~\cite{Uszko2025Silver}   &                                                \\
                    & Silver 5,5'-azotetrazolate             &                                                \\
Au                  & Fulminating gold~\cite{Uszko2024Gold}    & Au                                             \\
Pt                  & Potassium tetrachloroplatinate(II)–sodium azide reaction mixture & Platinum oxides                                             \\
Pd                  & Palladium 5,5'-azotetrazolate         & Pd                                             \\ \hline
\multicolumn{3}{l}{$^*$ Sample prepared in nitrogen atmosphere. All other samples were prepared in air.}                   
\end{tabular}
\end{table*}

\subsection{Lead azide and lead picrate}
\begin{figure}[htbp]
    \centering
    \begin{minipage}[t]{0.20\textwidth}
        \centering
        \includegraphics[width=\linewidth]{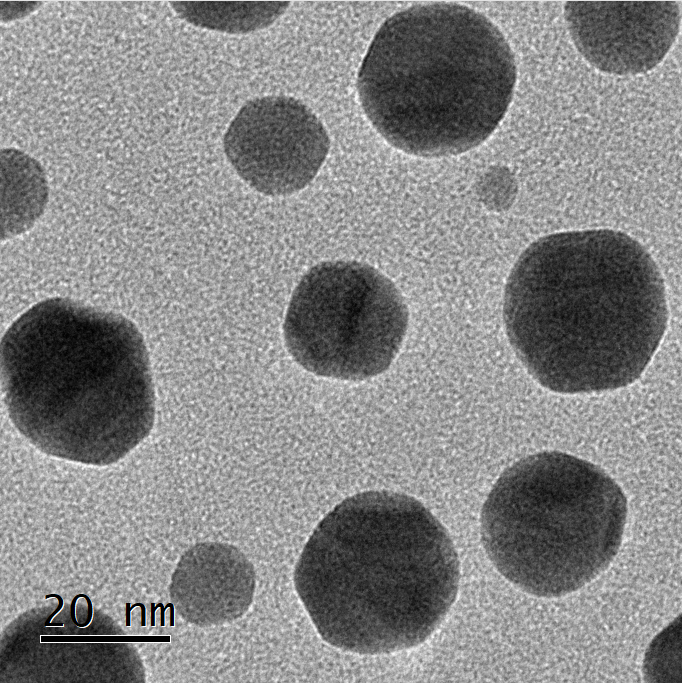}\\
        \textbf{(a)}
    \end{minipage}
    \hfill
    \begin{minipage}[t]{0.20\textwidth}
        \centering
        \includegraphics[width=\linewidth]{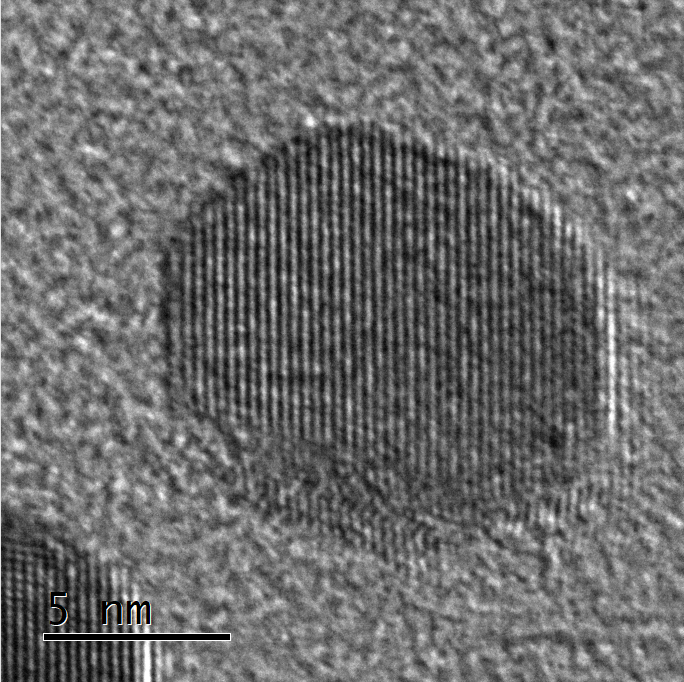}\\
        \textbf{(b)}
    \end{minipage}

    \vspace{0.5em}

    \begin{minipage}[t]{0.20\textwidth}
        \centering
        \includegraphics[width=\linewidth]{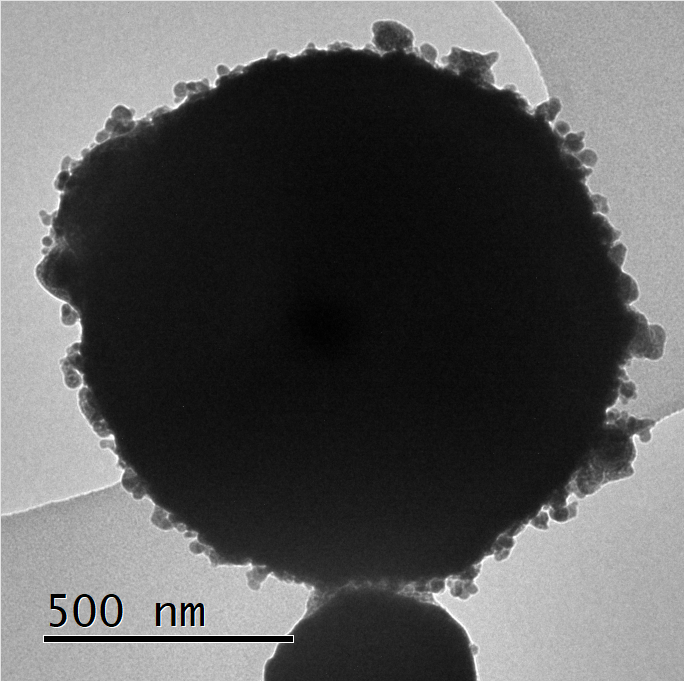}\\
        \textbf{(c)}
    \end{minipage}
    \hfill
    \begin{minipage}[t]{0.20\textwidth}
        \centering
        \includegraphics[width=\linewidth]{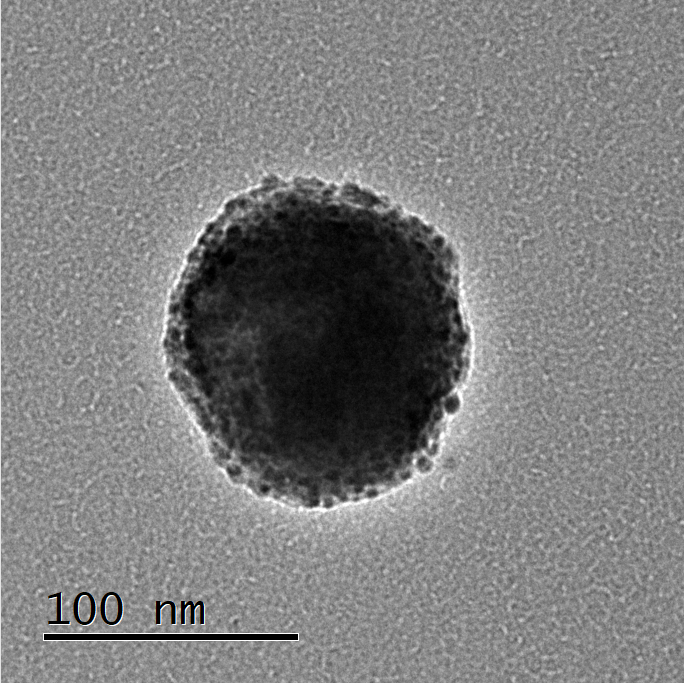}\\
        \textbf{(d)}
    \end{minipage}
    \caption{TEM micrographs of particles produced by detonation of (a--c) lead azide and (d) lead picrate.}
    \label{fig:SleadTEM}
\end{figure}

The detonation of lead-containing energetic precursors produced highly crystalline particles with a broad size distribution. Representative particles from lead azide and lead picrate are shown in \autoref{fig:SleadTEM}.

\begin{figure}[htbp]
    \centering
    \begin{minipage}[t]{0.20\textwidth}
        \centering
        \includegraphics[width=\linewidth]{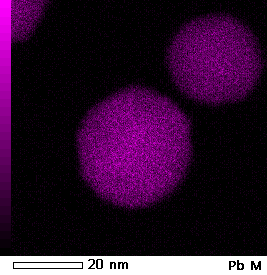}\\
        \textbf{(a)}
    \end{minipage}
    \hfill
    \begin{minipage}[t]{0.20\textwidth}
        \centering
        \includegraphics[width=\linewidth]{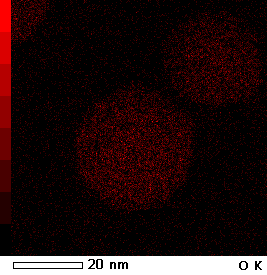}\\
        \textbf{(b)}
    \end{minipage}
    \vspace{0.5em}
    \begin{minipage}[t]{0.20\textwidth}
        \centering
        \includegraphics[width=\linewidth]{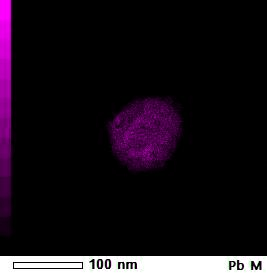}\\
        \textbf{(c)}
    \end{minipage}
    \hfill
    \begin{minipage}[t]{0.20\textwidth}
        \centering
        \includegraphics[width=\linewidth]{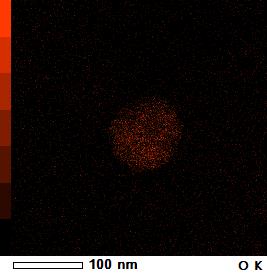}\\
        \textbf{(d)}
    \end{minipage}
    \caption{EDX maps showing Pb and O distributions for particles produced from (a,b) lead azide and (c,d) lead picrate.}
    \label{fig:SleadEDX}
\end{figure}

EDX mapping in~\autoref{fig:SleadEDX} confirmed co-localised lead and oxygen signals. 

Examination of a large number of particles produced by the detonation of lead-based energetic precursors revealed a strikingly consistent structural feature: a large proportion of crystalline particles for which lattice fringes could be resolved exhibited a \textit{d}-spacing of approximately 2.85~\AA. Together with the co-localised Pb and O signals observed by EDX, this recurring lattice spacing supports the formation of a lead oxide during detonation. Recent \textit{in situ} studies by Kinnibrugh et al. identified metastable Pb$_3$O$_5$ and Pb$_2$O$_3$ intermediates during the thermal decomposition of PbO$_2$ to PbO.~\cite{Kinnibrugh2024} Earlier work by Bystr\"om similarly established the formation of intermediate lead oxides during the decomposition of lead peroxide.~\cite{Bystrom1945Decomposition} The observed 2.85~\AA{} spacing is compatible with reflections from both Pb$_3$O$_4$ (ICSD \#4106) and Pb$_3$O$_5$ (ICSD \#36294). The combined observations are therefore consistent with the recovered particles representing material along the PbO$_2$ decomposition pathway towards the more stable Pb$_3$O$_4$, although the single observed lattice spacing does not permit these two phases to be distinguished unambiguously.
\subsection{Bismuth-containing energetic precursor}

\begin{figure}[htbp]
    \centering
    \begin{minipage}[t]{0.20\textwidth}
        \centering
        \includegraphics[width=\linewidth]{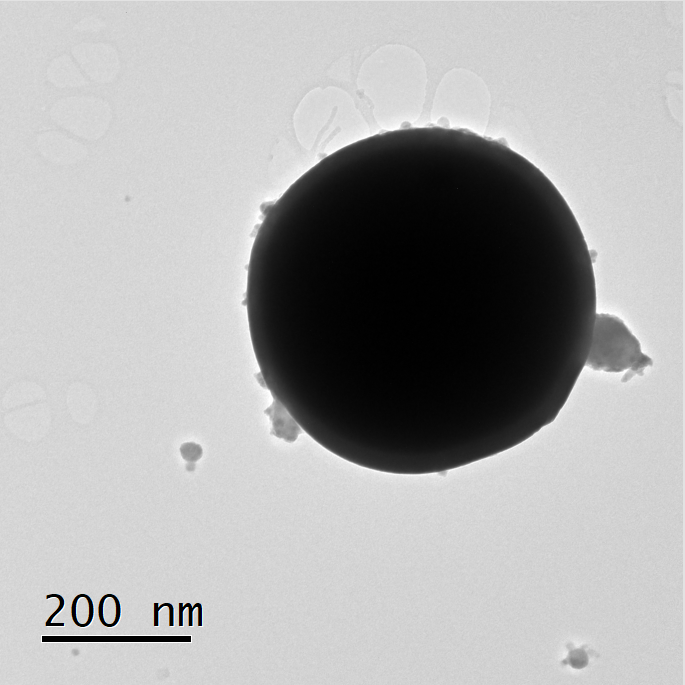}\\
        \textbf{(a)}
    \end{minipage}
    \hfill
    \begin{minipage}[t]{0.20\textwidth}
        \centering
        \includegraphics[width=\linewidth]{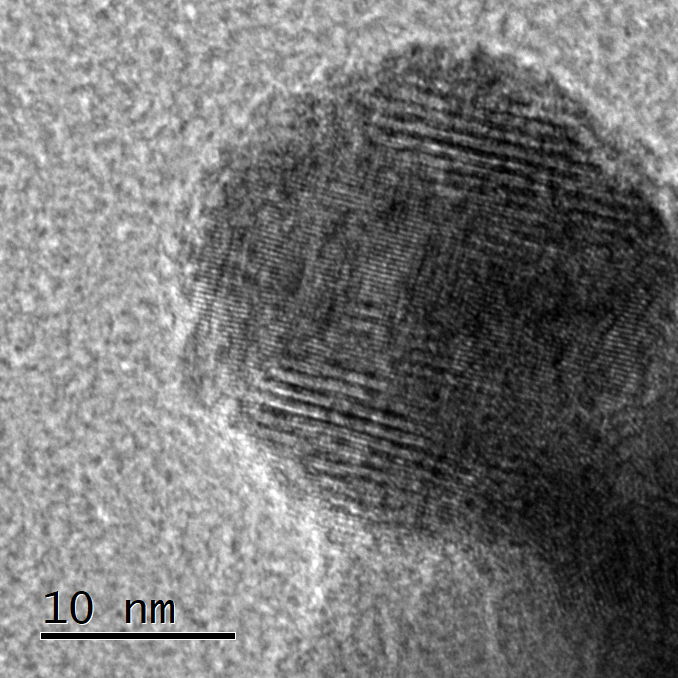}\\
        \textbf{(b)}
    \end{minipage}
    \caption{Representative TEM micrographs of particles produced by detonation of basic bismuth 5,5'-azotetrazolate.}
    \label{fig:SBiTEM}
\end{figure}

\begin{figure}[htbp]
    \centering
    \begin{minipage}[t]{0.2\textwidth}
        \centering
        \includegraphics[width=\linewidth]{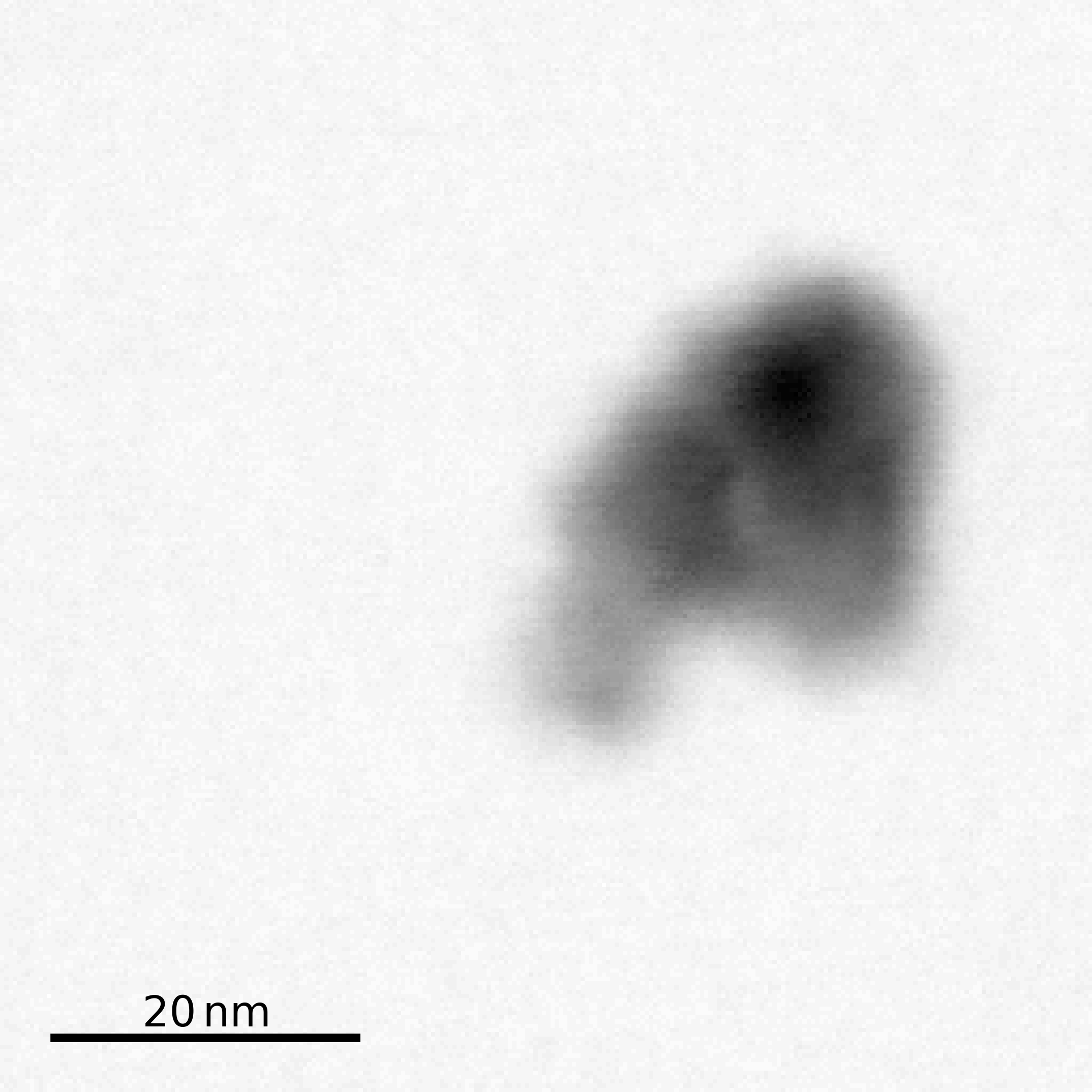}\\
        \textbf{(a)}
    \end{minipage}
    \hspace{0.04\textwidth}
    \begin{minipage}[t]{0.2\textwidth}
        \centering
        \includegraphics[width=\linewidth]{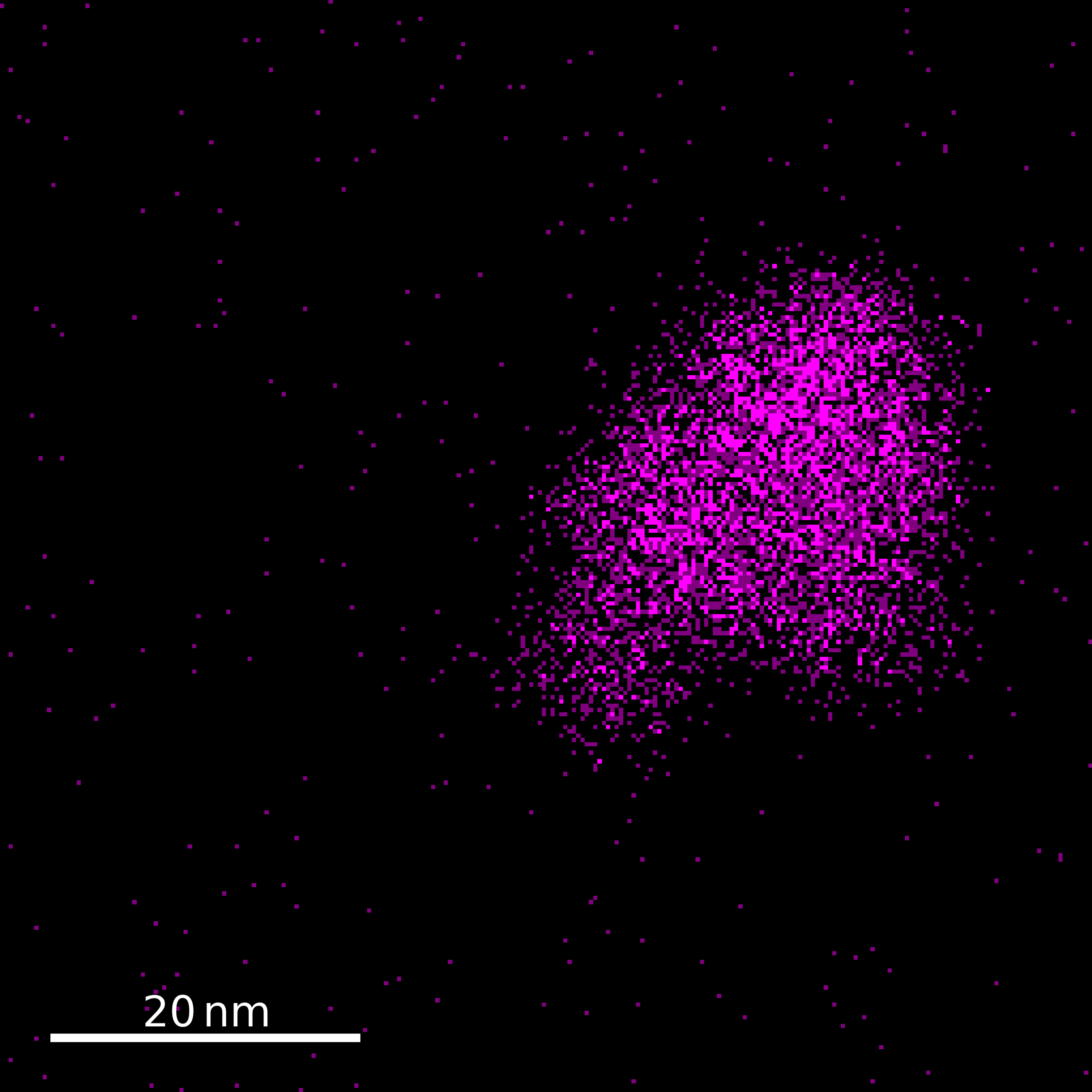}\\
        \textbf{(b)}
    \end{minipage}
    \hspace{0.04\textwidth}
    \begin{minipage}[t]{0.2\textwidth}
        \centering
        \includegraphics[width=\linewidth]{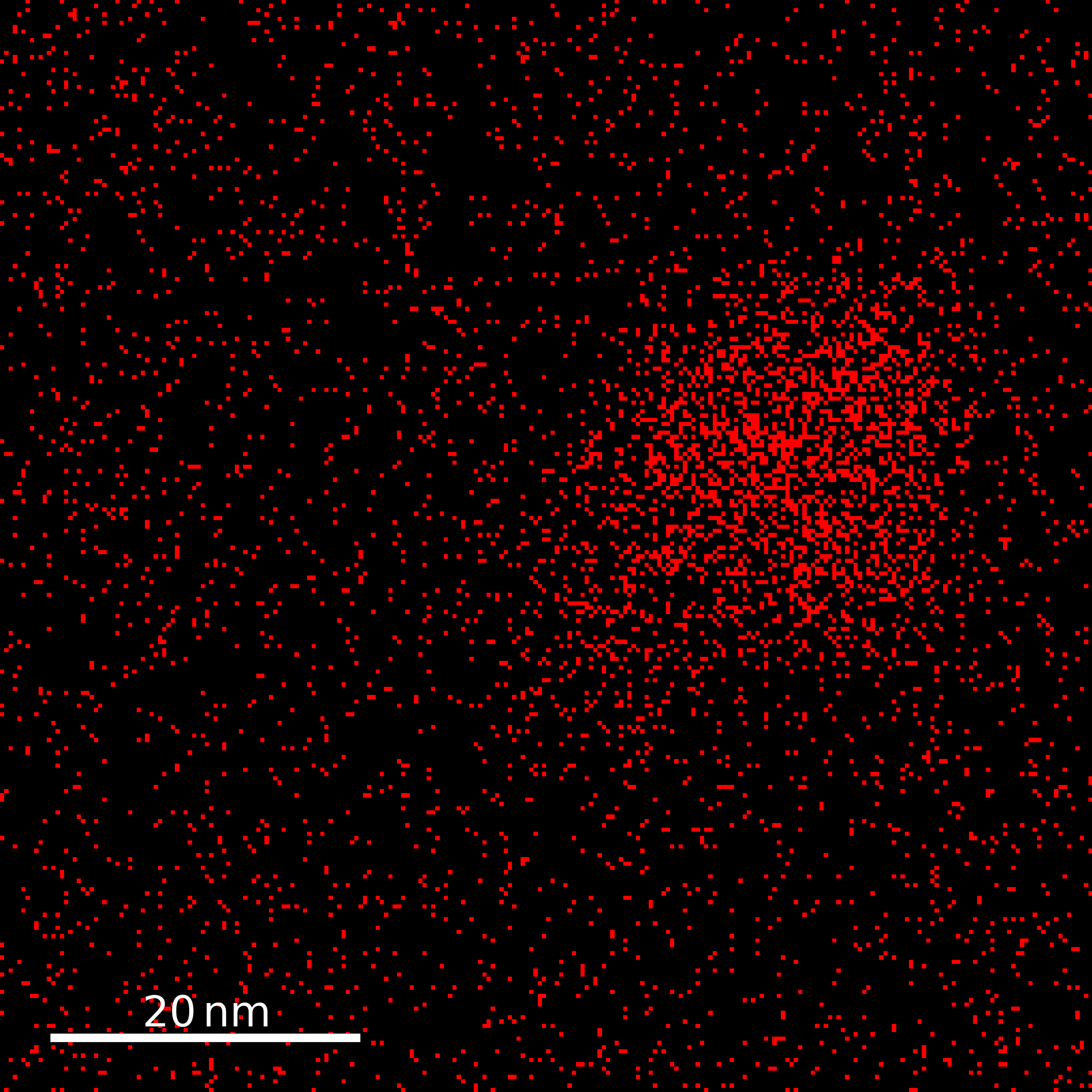}\\
        \textbf{(c)}
    \end{minipage}
    \caption{A particle cluster (a) electron micrograph and EDX maps for (b) Bi and (c) O.}
    \label{fig:SBiEDX}
\end{figure}

TEM micrographs of particles produced by detonation of basic bismuth 5,5'-azotetrazolate in~\autoref{fig:SBiTEM} showed highly crystalline particles spanning tens to hundreds of nanometres. Energy-dispersive X-ray spectroscopy (EDX) analysis of a cluster of particles in~\autoref{fig:SBiEDX} confirmed the presence of bismuth and oxygen.

\begin{table}[htbp]
    \caption{Local lattice-fringe measurements for the bismuth-containing sample.}
    \label{tab:SBiLattice}
    \centering
    \begin{tabular}{c|cc}
        Measured & \multicolumn{2}{c}{Bi$_2$O$_3$, ICSD \#15072}\\
        \hline
        \textit{d}-spacing/\AA & reference \textit{d}-spacing/\AA & lattice plane\\
        \hline
        2.76 & 2.75 & $(2\,1\,\bar{1})$\\
        2.73 & 2.71 & $(1\,2\,\bar{2})$\\
        2.69 & 2.69 & (2 0 0)\\
    \end{tabular}
\end{table}

The measured local lattice spacings in~\autoref{tab:SBiLattice} are consistent with Bi$_2$O$_3$.
\subsection{Thallium-containing energetic precursor}
\begin{figure}[htbp]
    \centering
    \begin{minipage}[t]{0.35\textwidth}
        \centering
        \includegraphics[width=\linewidth]{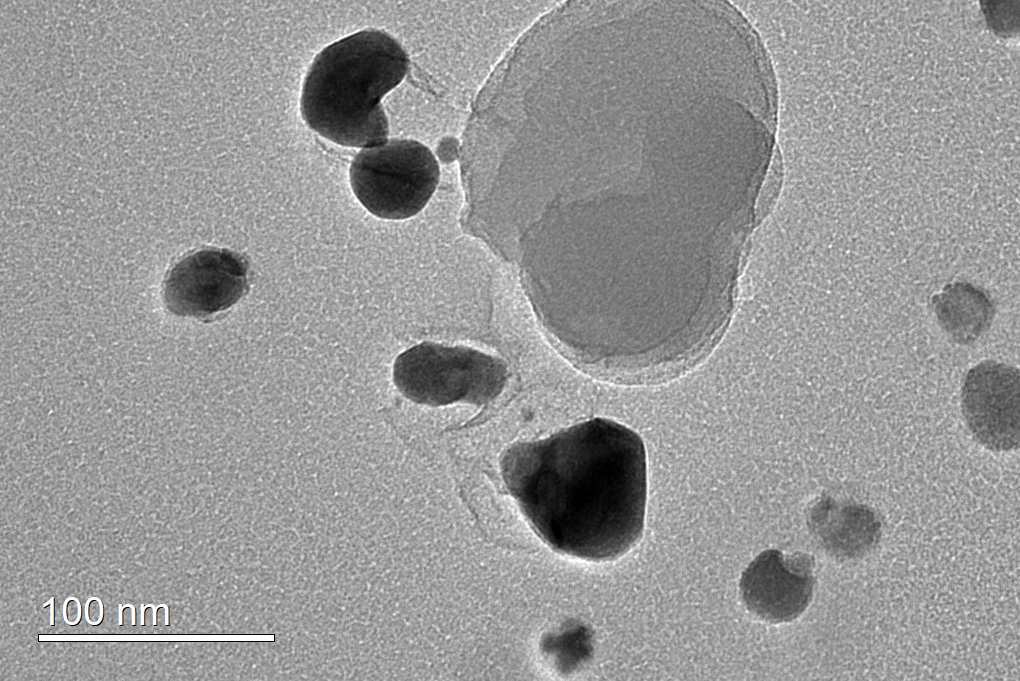}\\
        \textbf{(a)}
    \end{minipage}
    \hfill
    \begin{minipage}[t]{0.35\textwidth}
        \centering
        \includegraphics[width=\linewidth]{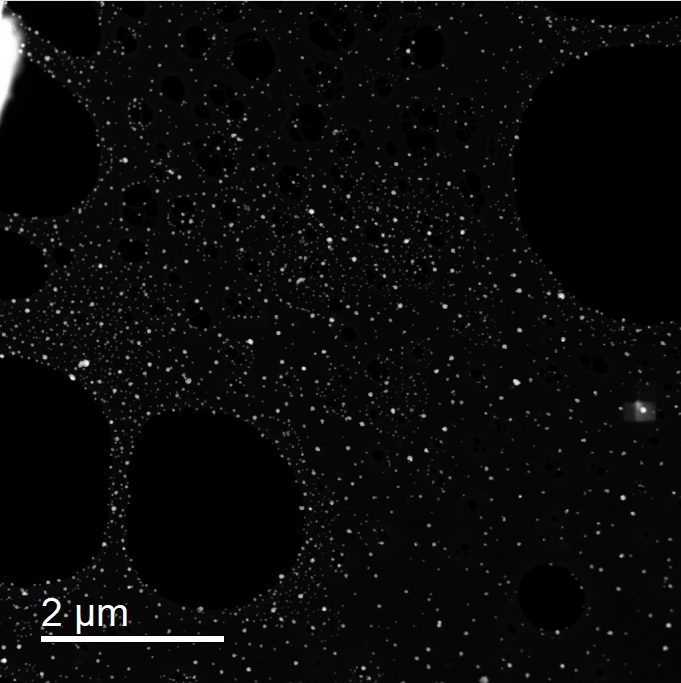}\\
        \textbf{(b)}
    \end{minipage}
    \caption{Bright-field and dark-field TEM images of particles produced by detonation of thallium azide.}
    \label{fig:STlTEM}
\end{figure}

\begin{figure}[htbp]
    \centering
    \begin{minipage}[t]{0.2\textwidth}
        \centering
        \includegraphics[width=\linewidth]{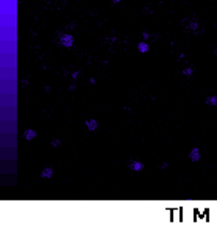}\\
        \textbf{(a)}
    \end{minipage}
    \hspace{0.04\textwidth}
    \begin{minipage}[t]{0.2\textwidth}
        \centering
        \includegraphics[width=\linewidth]{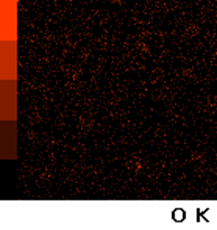}\\
        \textbf{(b)}
    \end{minipage}
    \caption{EDX maps showing (a) Tl and (b) O.}
    \label{fig:STlEDX}
\end{figure}

TEM micrographs of particles produced by detonation of thallium azide in~\autoref{fig:STlTEM} showed highly crystalline particles spanning tens to hundreds of nanometres. The EDX analysis of particles in~\autoref{fig:STlEDX} confirmed the presence of oxygen and thallium.

\begin{table*}[htbp]
    \caption{Local lattice-fringe measurements for the thallium-containing sample.}
    \label{tab:STlLattice}
    \centering
    \begin{tabular}{c|cc|cc}
        Measured & \multicolumn{2}{c}{(Tl$_2$O$_3$)$_{0.89}$, ICSD \#54214} & \multicolumn{2}{c}{Tl$_4$O$_3$, ICSD \#23478}\\
        \hline
        \textit{d}-spacing/\AA & reference \textit{d}-spacing/\AA & lattice plane& reference \textit{d}-spacing/\AA & lattice plane\\
        \hline
        3.89 & 3.95 & (2 0 0)&3.80&(10$\bar{2}$)\\
        3.86 & 3.95 & (2 0 0)&3.80&(10$\bar{2}$)\\
    \end{tabular}
\end{table*}
The measured lattice spacings in~\autoref{tab:STlLattice} of 3.89 and 3.86 \AA~are consistent with a thallium oxide phase. Although these spacings are compatible with the $(200)$ reflection of the examined Tl$_2$O$_3$-derived reference structure, other thallium oxides exhibit reflections in the same region. The local lattice-fringe measurements therefore support the formation of Tl--O-containing crystalline particles but do not permit an unambiguous assignment of oxide stoichiometry.

\subsection{Manganese-containing energetic precursor}
\begin{figure}[htbp]
    \centering
    \begin{minipage}[t]{0.45\textwidth}
        \centering
        \includegraphics[width=\linewidth]{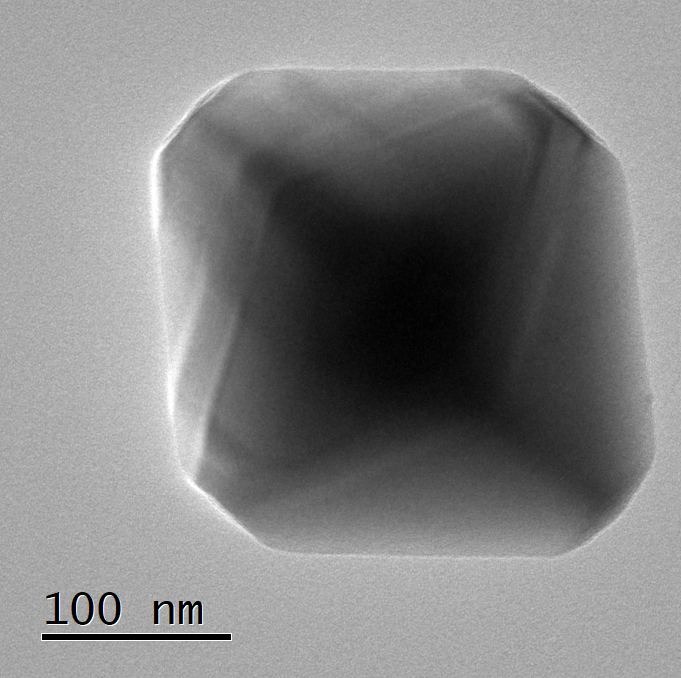}\\
        \textbf{(a)}
    \end{minipage}
    \hfill
    \begin{minipage}[t]{0.45\textwidth}
        \centering
        \includegraphics[width=\linewidth]{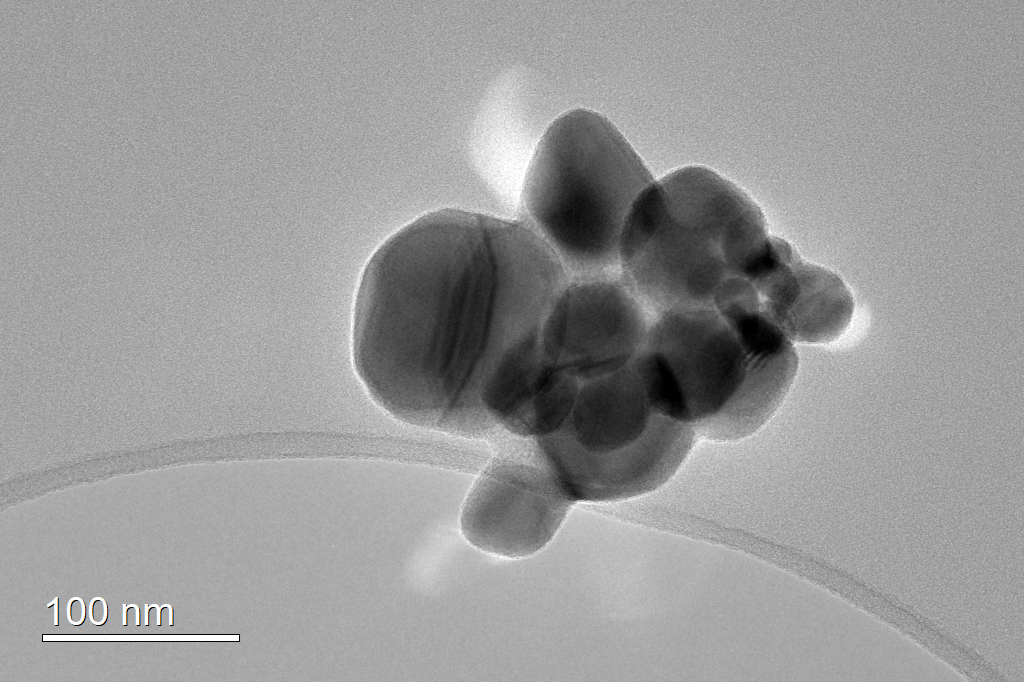}\\
        \textbf{(b)}
    \end{minipage}
    \caption{Representative TEM images of particles produced by detonation of manganese tris(carbohydrazide) perchlorate.}
    \label{fig:SMnTEM}
\end{figure}

\begin{figure}[htbp]
    \centering
    \begin{minipage}[t]{0.2\textwidth}
        \centering
        \includegraphics[width=\linewidth]{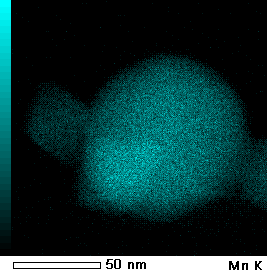}\\
        \textbf{(a)}
    \end{minipage}
    \hspace{0.04\textwidth}
    \begin{minipage}[t]{0.2\textwidth}
        \centering
        \includegraphics[width=\linewidth]{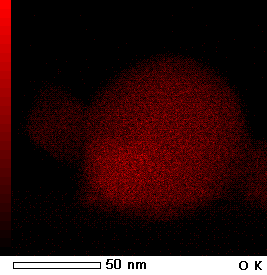}\\
        \textbf{(b)}
    \end{minipage}
    \caption{EDX maps showing (a) Mn and (b) O.}
    \label{fig:SMnEDX}
\end{figure}

TEM micrographs of particles produced by detonating manganese tris(carbohydrazide) perchlorate in~\autoref{fig:SMnTEM} showed highly crystalline octahedral particles.  The EDX analysis of particles in~\autoref{fig:SMnEDX} confirmed the presence of oxygen and manganese.

\begin{table}[htbp]
    \caption{Local lattice-fringe measurement for the manganese-containing sample.}
    \label{tab:SMnLattice}
    \centering
    \begin{tabular}{c|cc}
        Measured & \multicolumn{2}{c}{Mn$_3$O$_4$, ICSD \#33327}\\
        \hline
        \textit{d}-spacing/\AA & reference \textit{d}-spacing/\AA & lattice plane\\
        \hline
        2.92 & 2.89 & (2 0 0)\\
    \end{tabular}
\end{table}
The measured local spacing in\autoref{tab:SMnLattice} is consistent with Mn$_3$O$_4$.

\subsection{Iron-containing energetic precursor}
\begin{figure}[htbp]
    \centering
    \begin{minipage}[t]{0.48\textwidth}
        \centering
        \includegraphics[width=\linewidth]{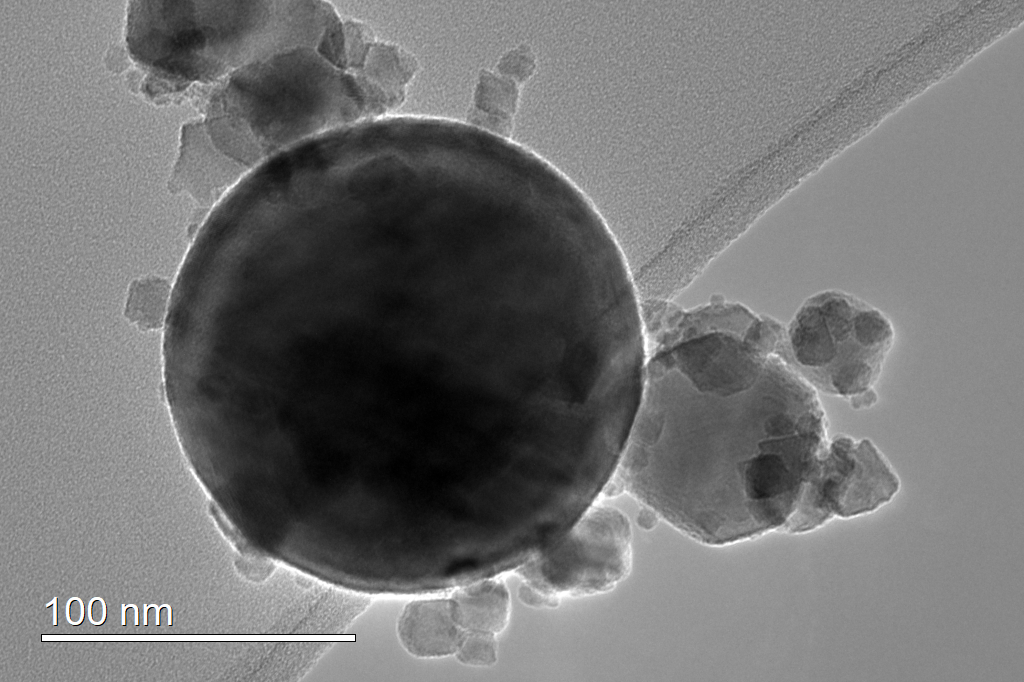}\\
        \textbf{(a)}
    \end{minipage}
    \hfill
    \begin{minipage}[t]{0.48\textwidth}
        \centering
        \includegraphics[width=\linewidth]{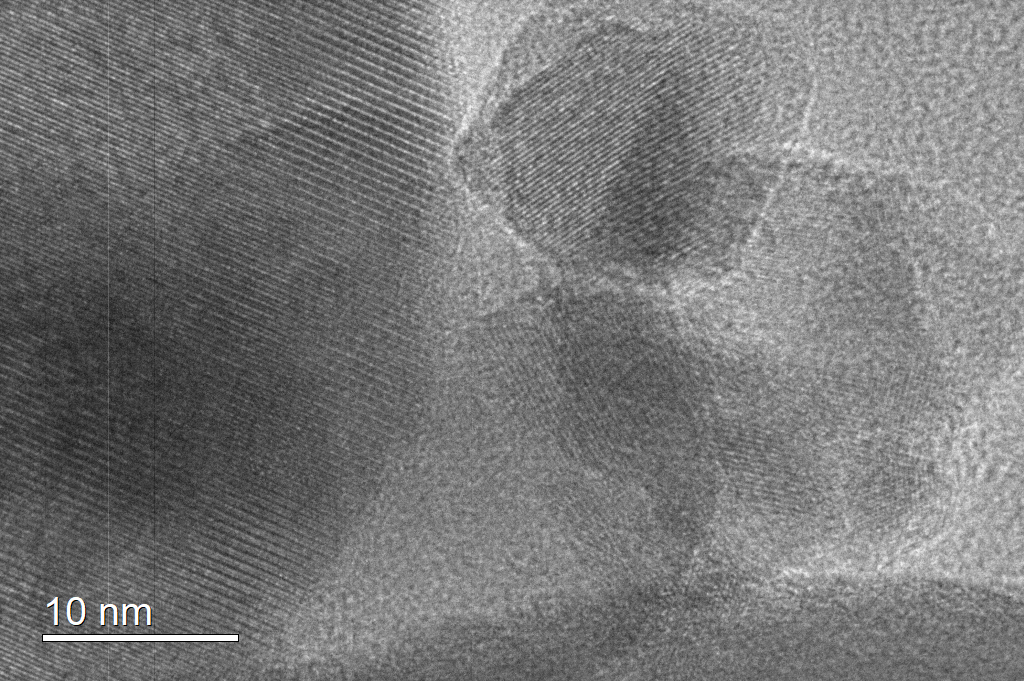}\\
        \textbf{(b)}
    \end{minipage}
    \caption{Representative TEM and HRTEM images of particles produced by detonation of FeCP.}
    \label{fig:SFeTEM}
\end{figure}

\begin{figure}[htbp]
    \centering
    \begin{minipage}[t]{0.2\textwidth}
        \centering
        \includegraphics[width=\linewidth]{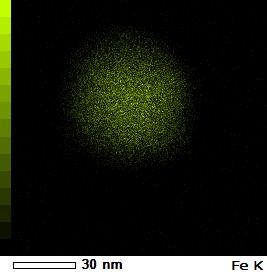}\\
        \textbf{(a)}
    \end{minipage}
    \hspace{0.04\textwidth}
    \begin{minipage}[t]{0.2\textwidth}
        \centering
        \includegraphics[width=\linewidth]{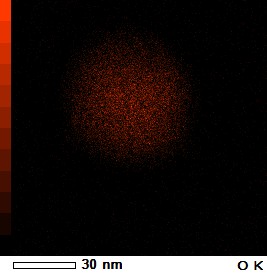}\\
        \textbf{(b)}
    \end{minipage}
    \caption{EDX maps showing (a) Fe and (b) O of particles.}
    \label{fig:SFeEDX}
\end{figure}

TEM micrographs of particles produced by detonating iron tris(carbohydrazide) perchlorate (FeCP) in~\autoref{fig:SFeTEM} showed highly crystalline clustered particles.  The EDX analysis of particles in~\autoref{fig:SFeEDX} confirmed the presence of oxygen and iron.

\begin{table}[htbp]
    \caption{Local lattice-fringe measurement from the iron-containing sample.}
    \label{tab:SFeLattice}
    \centering
    \begin{tabular}{c|cc}
        Measured & \multicolumn{2}{c}{Fe$_4$O$_5$, ICSD \#47905}\\
        \hline
        \textit{d}-spacing/\AA & reference \textit{d}-spacing/\AA & lattice plane\\
        \hline
        2.55 & 2.54 & (112)\\
    \end{tabular}
\end{table}

\begin{figure}[htbp]
    \centering
    \includegraphics[width=.8\linewidth]{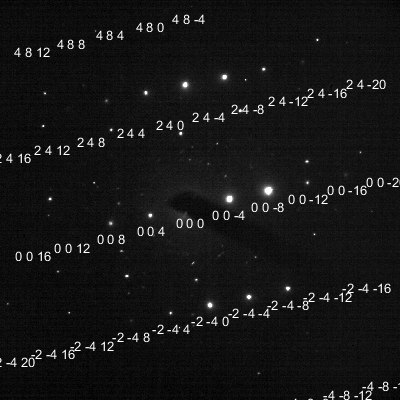}
    \caption{Electron diffraction pattern obtained from an iron-containing particle.}
    \label{fig:SFeED}
\end{figure}

The iron-containing product is a mixture of crystalline iron oxides. A local spacing in~\autoref{tab:SFeLattice} is compatible with Fe$_4$O$_5$, but the electron diffraction pattern,~\autoref{fig:SFeED}, contains contributions from multiple crystalline phases and does not permit an unambiguous phase assignment. The Fe$_4$O$_5$ comparison is therefore retained as a local observation rather than a claim of a pure high-pressure phase.
\subsection{Copper-containing energetic precursors}

\subsubsection{Tetraamminecopper perchlorate}
\begin{figure}[htbp]
    \centering
    \begin{minipage}[t]{0.48\textwidth}
        \centering
        \includegraphics[width=\linewidth]{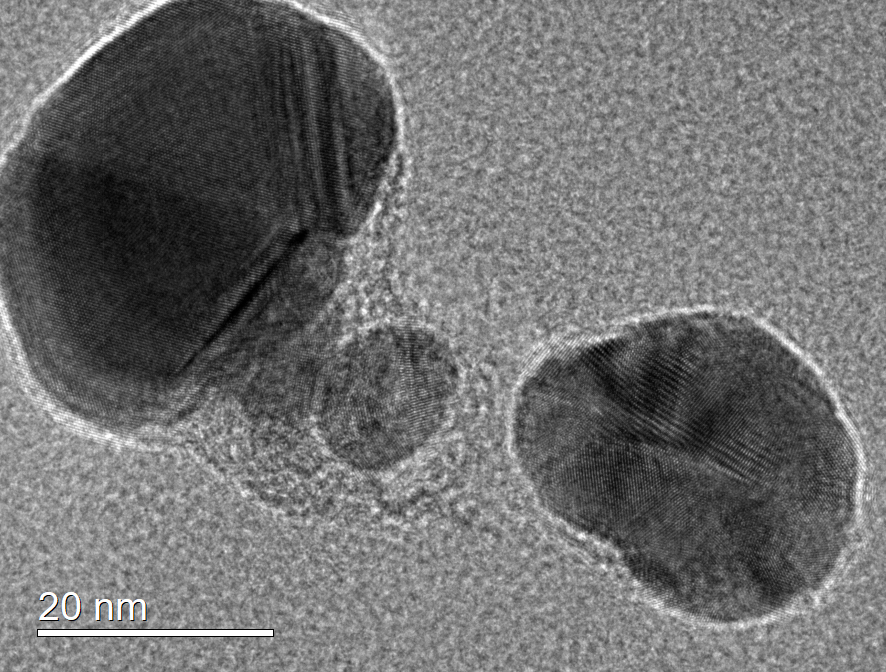}\\
        \textbf{(a)}
    \end{minipage}
    \hfill
    \begin{minipage}[t]{0.48\textwidth}
        \centering
        \includegraphics[width=\linewidth]{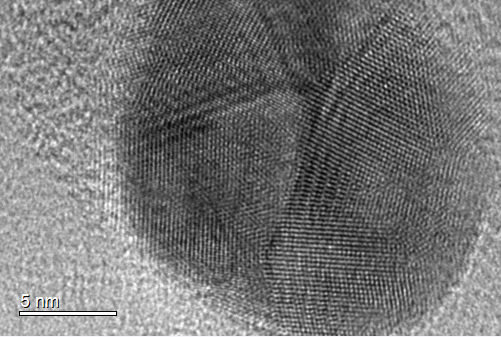}\\
        \textbf{(b)}
    \end{minipage}
    \caption{TEM and HRTEM images of copper-containing particles produced by detonation of TACP.}
    \label{fig:SCuA1TEM}
\end{figure}

\begin{figure}[htbp]
    \centering
    \begin{minipage}[t]{0.2\textwidth}
        \centering
        \includegraphics[width=\linewidth]{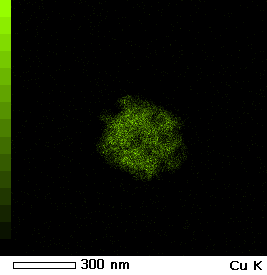}\\
        \textbf{(a)}
    \end{minipage}
    \hspace{0.04\textwidth}
    \begin{minipage}[t]{0.2\textwidth}
        \centering
        \includegraphics[width=\linewidth]{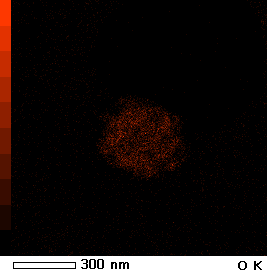}\\
        \textbf{(b)}
    \end{minipage}
    \caption{EDX maps showing (a) Cu and (b) O.}
    \label{fig:SCuA1EDX}
\end{figure}

TEM micrographs of particles produced by detonation of tetraamminecopper perchlorate (TACP) in~\autoref{fig:SCuA1TEM} showed examples of crystalline particles. EDX analysis in~\autoref{fig:SCuA1EDX} confirmed the presence of copper and oxygen.

\begin{table}[htbp]
    \caption{Local lattice-fringe measurements for the TACP product.}
    \label{tab:SCuA1Lattice}
    \centering
    \begin{tabular}{c|cc}
        Measured & \multicolumn{2}{c}{CuO, ICSD \#16025}\\
        \hline
        \textit{d}-spacing/\AA & reference \textit{d}-spacing/\AA & lattice plane\\
        \hline
        2.45 & 2.52 & (1 1 $\bar{1}$)\\
        2.31 & 2.31 & (2 0 0)\\
    \end{tabular}
\end{table}
The local spacings in~\autoref{tab:SCuA1Lattice} are consistent with CuO.

\subsubsection{Cuprous acetylide}
\begin{figure}[htbp]
    \centering
    \begin{minipage}[t]{0.2\textwidth}
        \centering
        \includegraphics[width=\linewidth]{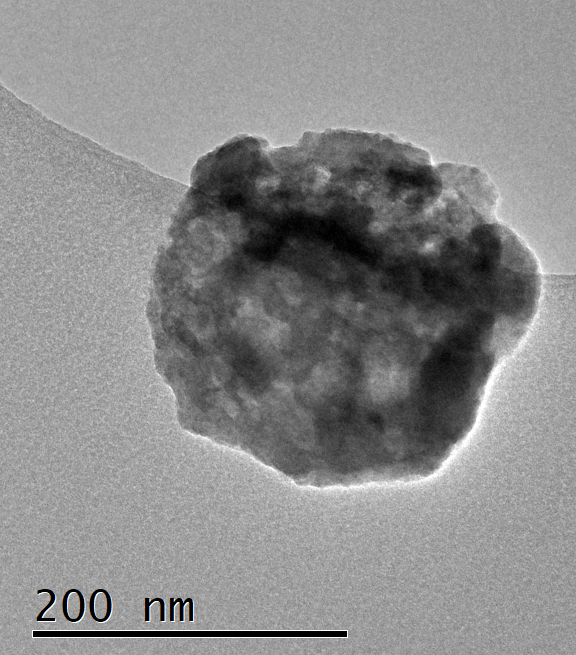}\\
        \textbf{(a)}
    \end{minipage}
    \hfill
    \begin{minipage}[t]{0.22\textwidth}
        \centering
        \includegraphics[width=\linewidth]{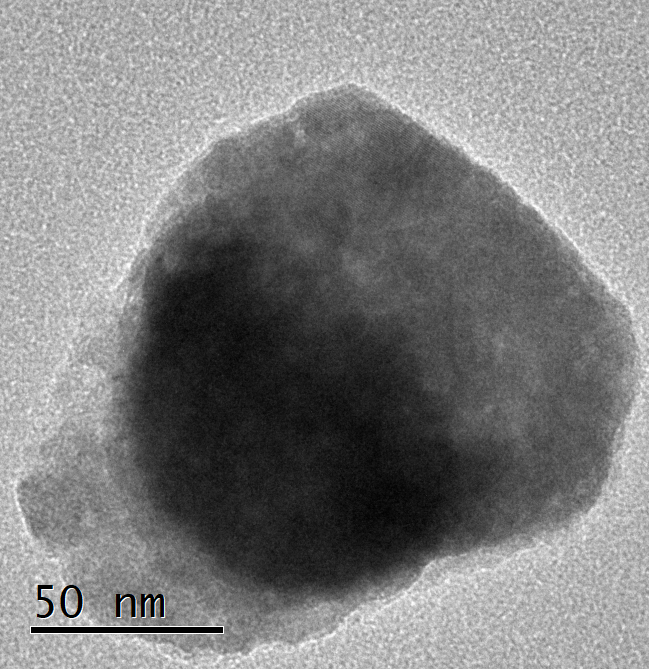}\\
        \textbf{(b)}
    \end{minipage}
    \caption{TEM micrographs of particles produced by detonation of cuprous acetylide.}
    \label{fig:SCuA5TEM}
\end{figure}

\begin{figure}[htbp]
    \centering
    \begin{minipage}[t]{0.2\textwidth}
        \centering
        \includegraphics[width=\linewidth]{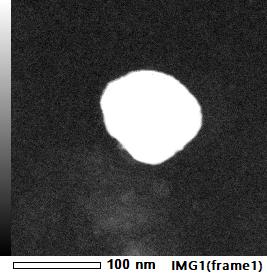}\\
        \textbf{(a)}
    \end{minipage}
    \hspace{0.04\textwidth}
    \begin{minipage}[t]{0.2\textwidth}
        \centering
        \includegraphics[width=\linewidth]{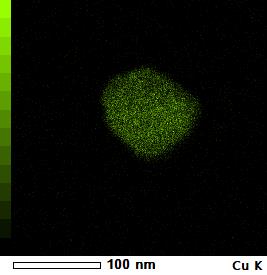}\\
        \textbf{(b)}
    \end{minipage}
    \hspace{0.04\textwidth}
    \begin{minipage}[t]{0.2\textwidth}
        \centering
        \includegraphics[width=\linewidth]{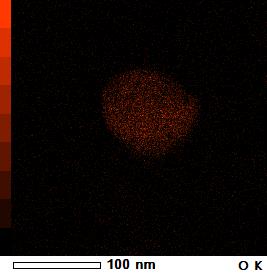}\\
        \textbf{(c)}
    \end{minipage}
    \caption{(a) Electron image and EDX maps for (b) Cu and (c) O from the cuprous acetylide product.}
    \label{fig:SCuA5EDX}
\end{figure}

TEM micrographs of particles produced by detonation of cuprous acetylide in~\autoref{fig:SCuA5TEM} showed examples of crystalline particle clusters. EDX analysis in~\autoref{fig:SCuA5EDX} confirmed the presence of copper and oxygen.

\begin{table}[htbp]
    \caption{Local lattice-fringe measurement from the cuprous acetylide product.}
    \label{tab:SCuA5Lattice}
    \centering
    \begin{tabular}{c|cc}
        Measured & \multicolumn{2}{c}{Cu$_4$O$_3$, ICSD \#77675}\\
        \hline
        \textit{d}-spacing/\AA & reference \textit{d}-spacing/\AA & lattice plane\\
        \hline
        2.48 & 2.47 & (0 0 4)\\
    \end{tabular}
\end{table}
The local lattice spacing measured  in~\autoref{tab:SCuA5Lattice} is compatible with Cu$_4$O$_3$. CuO is known to transform into Cu$_4$O$_3$, which suggests the initial formation of CuO as in the case of tetraamminecopper perchlorate. ~\cite{Morgan_1996,Long_1986}

\subsection{Zinc-containing energetic precursor}
\begin{figure}[htbp]
    \centering
    \begin{minipage}[t]{0.48\textwidth}
        \centering
        \includegraphics[width=\linewidth]{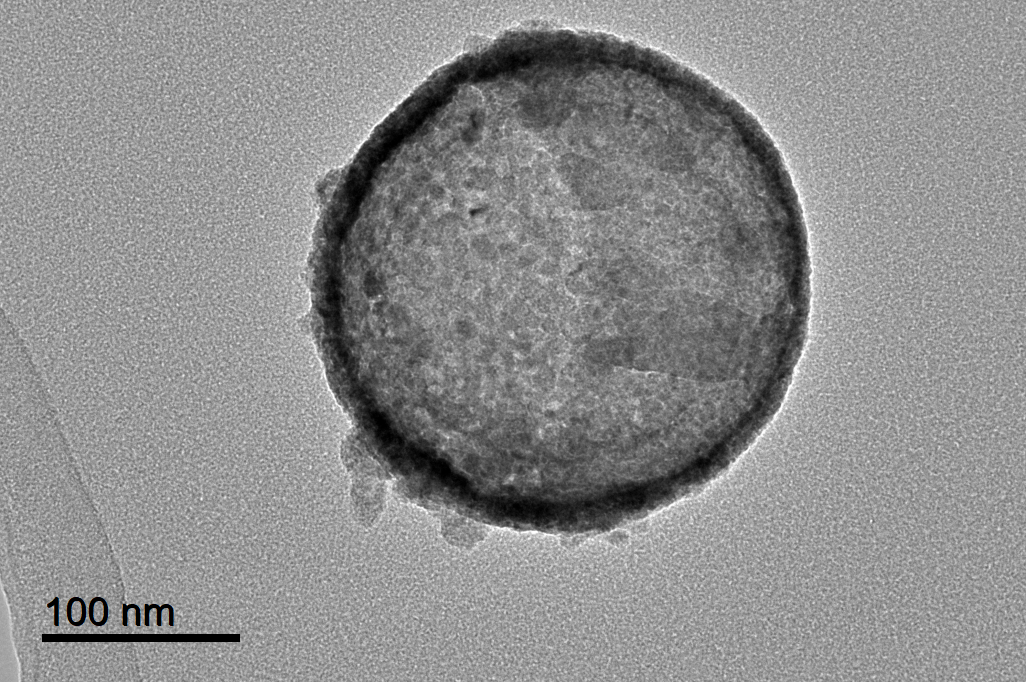}\\
        \textbf{(a)}
    \end{minipage}
    \hfill
    \begin{minipage}[t]{0.48\textwidth}
        \centering
        \includegraphics[width=\linewidth]{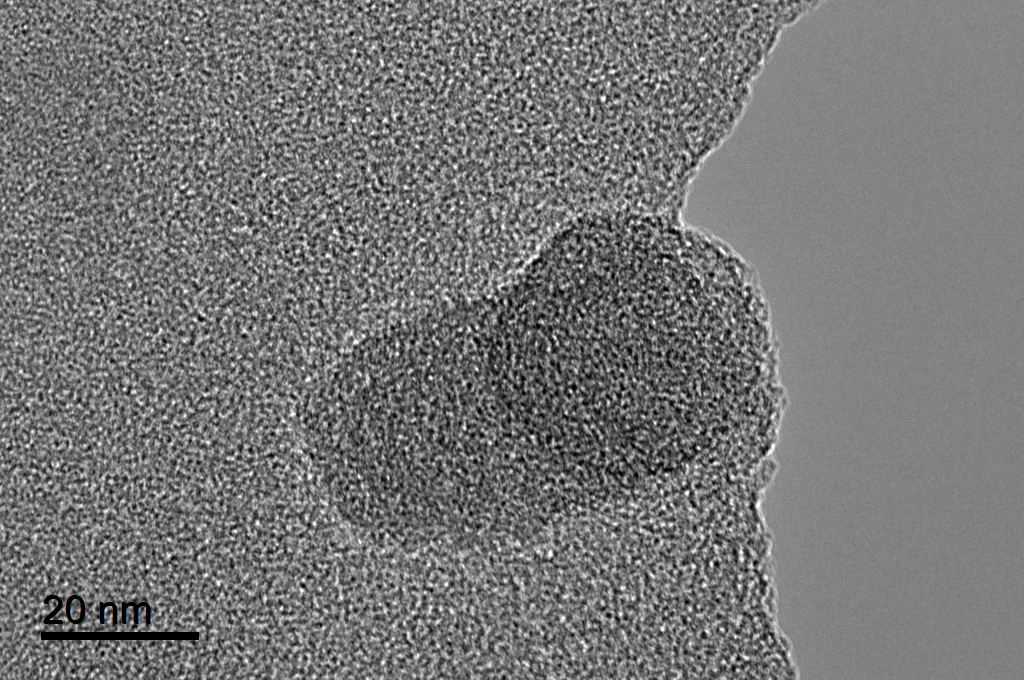}\\
        \textbf{(b)}
    \end{minipage}
    \caption{TEM and HRTEM images of particles produced by detonation of zinc 5,5'-azotetrazolate dihydrate.}
    \label{fig:SZnTEM}
\end{figure}

\begin{figure}[htbp]
    \centering
    \begin{minipage}[t]{0.2\textwidth}
        \centering
        \includegraphics[width=\linewidth]{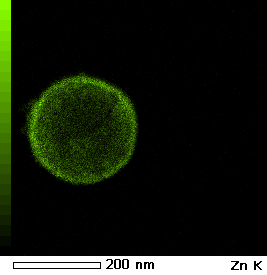}\\
        \textbf{(a)}
    \end{minipage}
    \hspace{0.04\textwidth}
    \begin{minipage}[t]{0.2\textwidth}
        \centering
        \includegraphics[width=\linewidth]{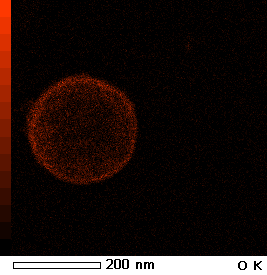}\\
        \textbf{(b)}
    \end{minipage}
    \caption{EDX maps showing (a) Zn and (b) O.}
    \label{fig:SZnEDX}
\end{figure}

TEM micrographs of particles produced by detonation of zinc 5,5'-azotetrazolate dihydrate in~\autoref{fig:SZnTEM} showed examples of crystalline particle clusters. EDX analysis in~\autoref{fig:SZnEDX} confirmed the presence of zinc and oxygen.

\begin{table}[htbp]
    \caption{Local lattice-fringe measurement for the zinc-containing sample.}
    \label{tab:SZnLattice}
    \centering
    \begin{tabular}{c|cc}
        Measured & \multicolumn{2}{c}{ZnO$_2$, ICSD \#38979}\\
        \hline
        \textit{d}-spacing/\AA & reference \textit{d}-spacing/\AA & lattice plane\\
        \hline
        2.44 & 2.45 & (2 0 0)\\
    \end{tabular}
\end{table}
The local spacing in~\autoref{tab:SZnLattice} is consistent with ZnO$_2$. This is a local assignment rather than a quantitative phase analysis of the complete product.

\subsection{Platinum-containing energetic precursor}
\begin{figure}[htbp]
    \centering
    \begin{minipage}[t]{0.28\textwidth}
        \centering
        \includegraphics[width=\linewidth]{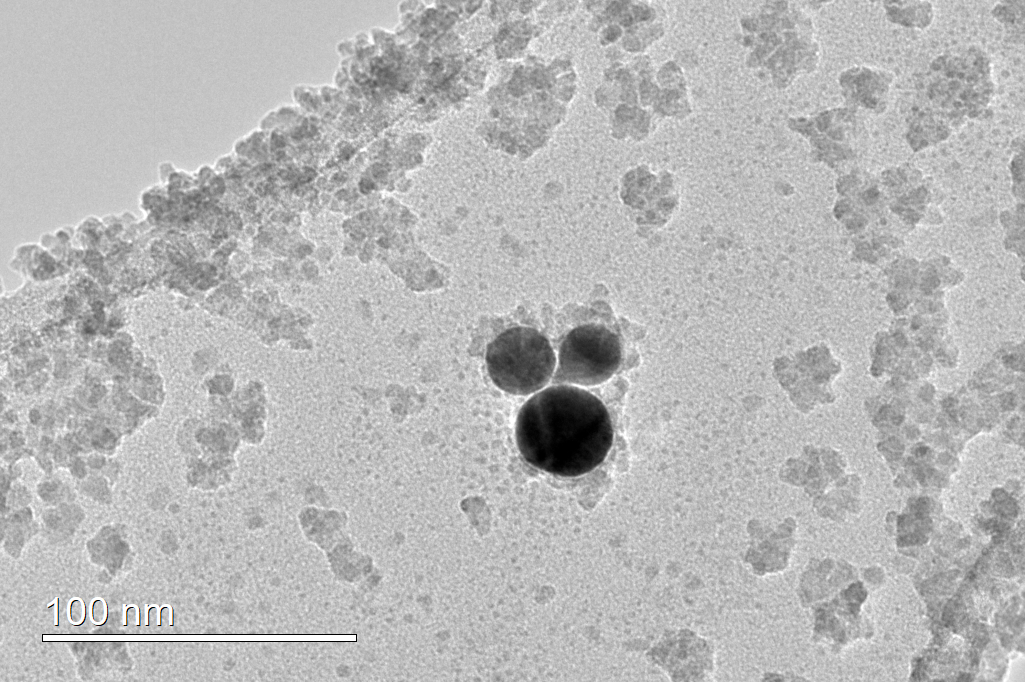}\\
        \textbf{(a)}
    \end{minipage}
    \hfill
    \begin{minipage}[t]{0.18\textwidth}
        \centering
        \includegraphics[width=\linewidth]{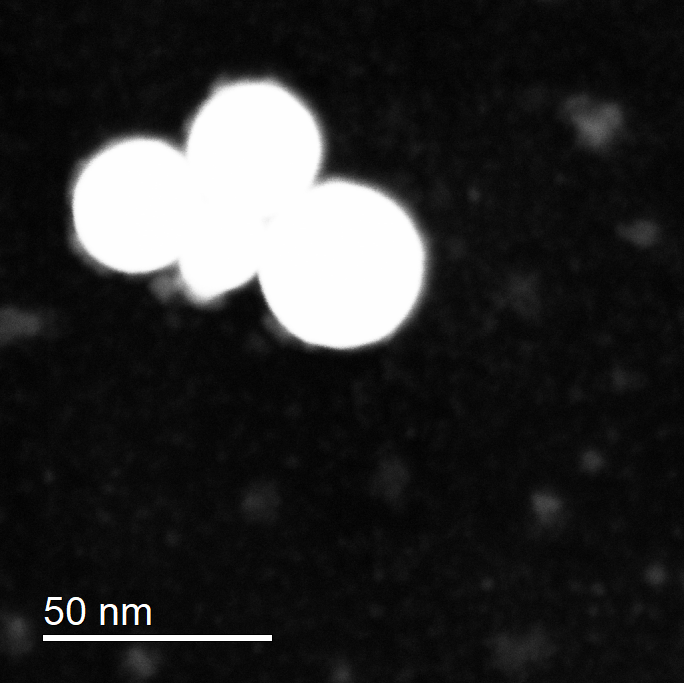}\\
        \textbf{(b)}
    \end{minipage}
    \caption{Bright-field and dark-field TEM images of particles produced by detonation of a platinum-containing precursor prepared from potassium tetrachloroplatinate(II) and sodium azide.}
    \label{fig:SPtTEM}
\end{figure}

\begin{figure}[htbp]
    \centering
    \begin{minipage}[t]{0.2\textwidth}
        \centering
        \includegraphics[width=\linewidth]{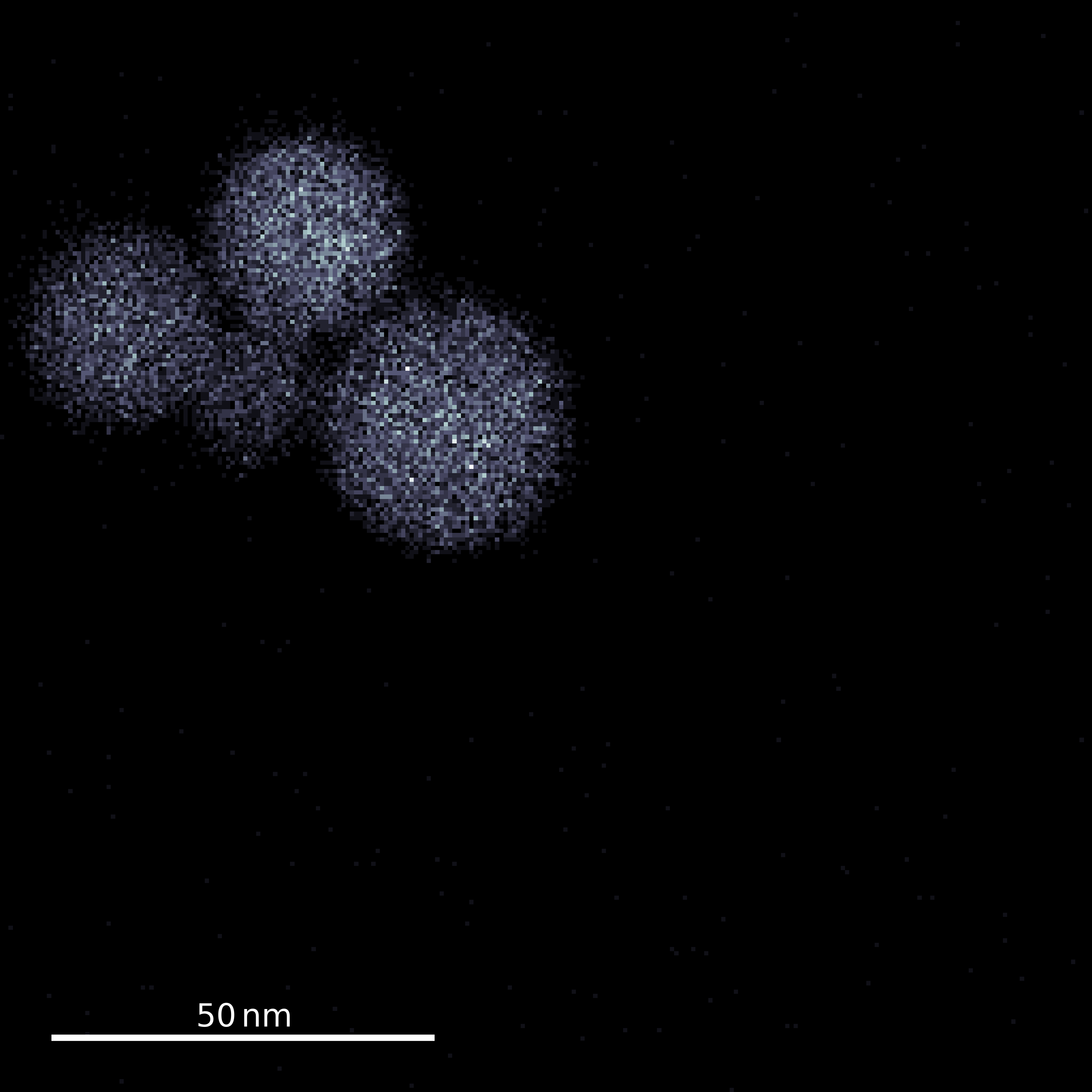}\\
        \textbf{(a)}
    \end{minipage}
    \hspace{0.04\textwidth}
    \begin{minipage}[t]{0.2\textwidth}
        \centering
        \includegraphics[width=\linewidth]{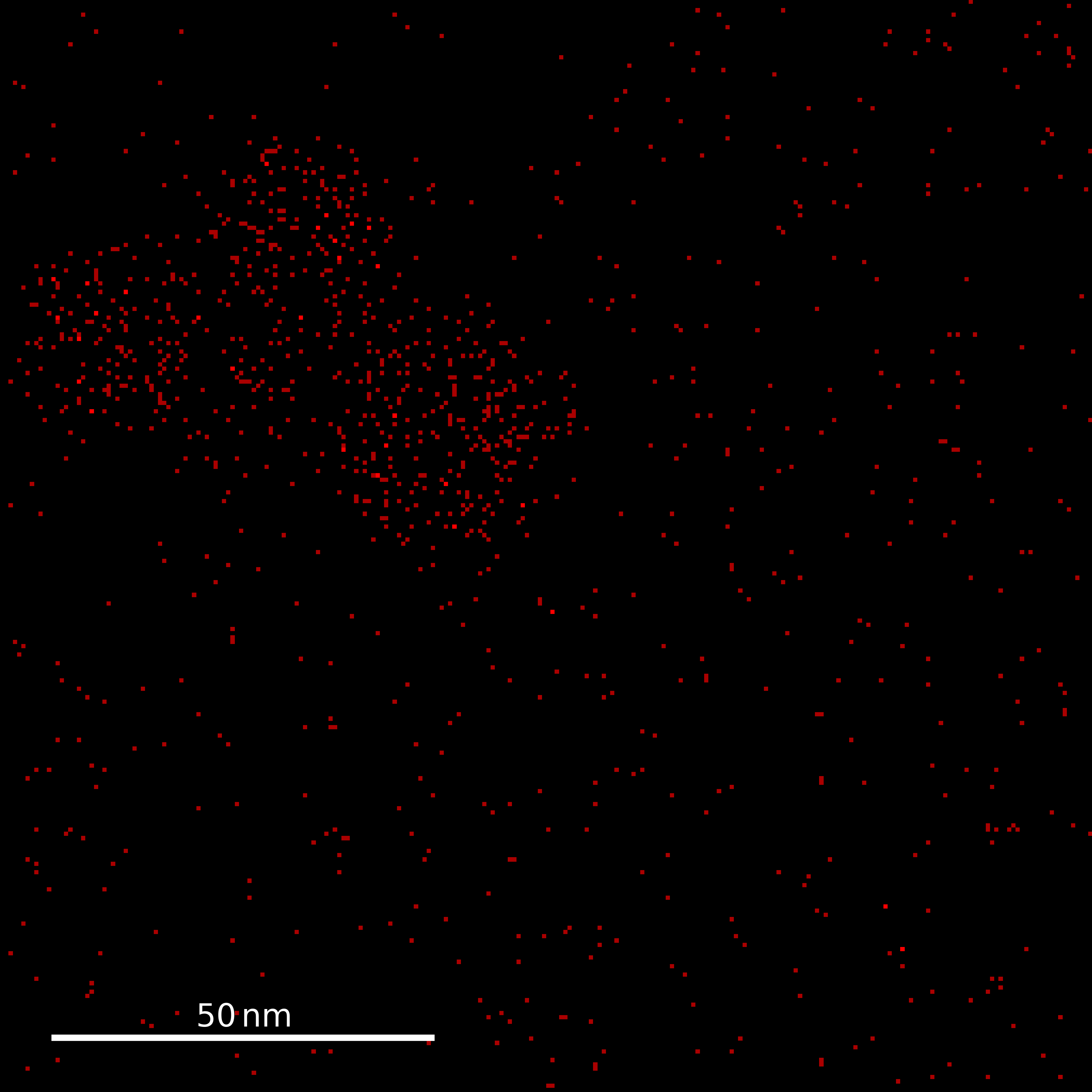}\\
        \textbf{(b)}
    \end{minipage}
    \caption{EDX maps for (a) Pt and (b) O.}
    \label{fig:SPtEDX}
\end{figure}

TEM micrographs of particles produced by detonation of a platinum-containing precursor prepared from potassium tetrachloroplatinate(II) and sodium azide are shown in~\autoref{fig:SPtTEM}. The EDX analysis of particles in~\autoref{fig:SPtEDX} confirmed the presence of oxygen and platinum in the particles.

\begin{table}[htbp]
    \caption{Local lattice-fringe measurement for the platinum-containing sample.}
    \label{tab:SPtLattice}
    \centering
    \begin{tabular}{c|cc}
        Measured & \multicolumn{2}{c}{PtO$_{2}$, ICSD \#251568}\\
        \hline
        \shortstack{\textit{d}-spacing\\(\AA)} &
        \shortstack{Reference spacing\\(\AA)} &
        \shortstack{Working plane\\assignment}\\
        \hline
        2.42 & 2.45 & (200)\\
        2.23 & 2.19 & (210)\\
    \end{tabular}
\end{table}

The measured local lattice spacings in~\autoref{tab:SPtLattice} are compatible with the examined high-pressure PtO$_2$ reference structure, although the available measurements do
not permit an unambiguous phase assignment.
\subsection{Nickel-containing energetic precursors}

Generally, the particles tend to be spherical or near-spherical with visible faceting, the two notable exceptions are particles produced by detonation of manganese- and nickel-based explosives. A range of nickel-containing explosives, including nickel tris(hydrazine) dinitrate (NHN), nickel tris(ethylenediamine) perchlorate (NiEP) and nickel tris(carbohydrazide) perchlorate  (NiCP) were tested.

\subsubsection{Nickel tris(hydrazine) dinitrate}

\begin{figure}[htbp]
    \centering
    \includegraphics[width=.60\linewidth]{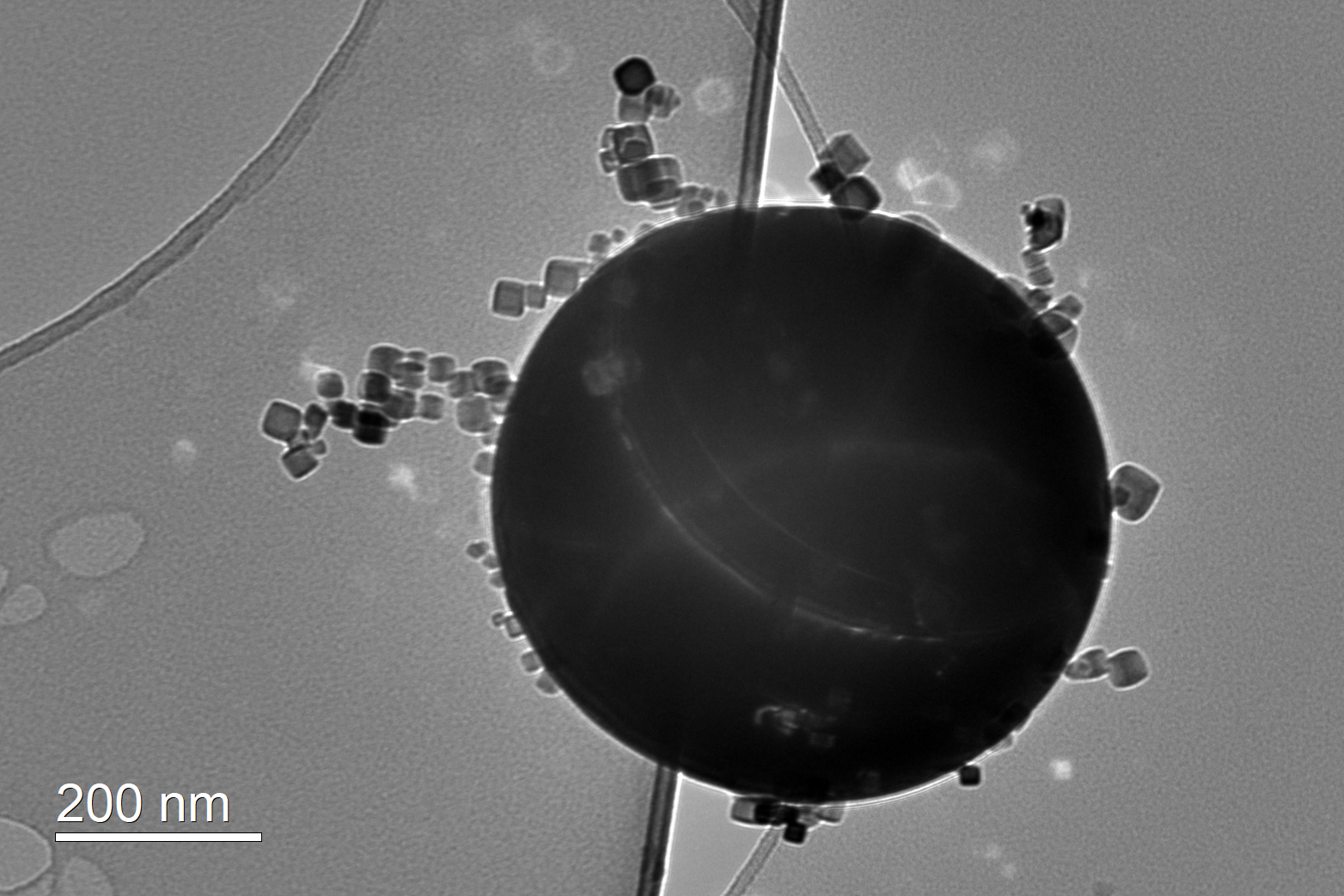}
    \caption{Representative TEM image of particles produced by detonation of NHN.}
    \label{fig:SNi12TEM}
\end{figure}

\begin{figure}[htbp]
    \centering
    \begin{minipage}[t]{0.2\textwidth}
        \centering
        \includegraphics[width=\linewidth]{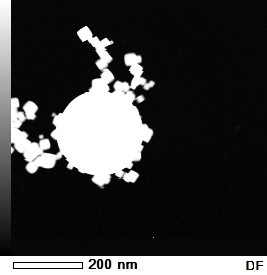}\\
        \textbf{(a)}
    \end{minipage}
    \hspace{0.04\textwidth}
    \begin{minipage}[t]{0.2\textwidth}
        \centering
        \includegraphics[width=\linewidth]{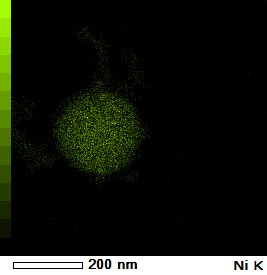}\\
        \textbf{(b)}
    \end{minipage}
    \hspace{0.04\textwidth}
    \begin{minipage}[t]{0.2\textwidth}
        \centering
        \includegraphics[width=\linewidth]{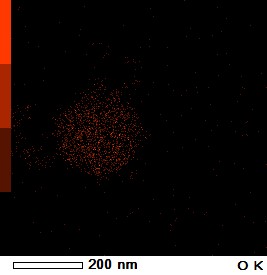}\\
        \textbf{(c)}
    \end{minipage}
    \caption{EDX map comparing the (a) dark-field image, (b) nickel content and (c) oxygen content of the particles prepared in the detonation of NHN.}
    \label{fig:A12EDX}
\end{figure}

TEM micrographs of particles produced by detonation of nickel tris(hydrazine) dinitrate (NHN) in~\autoref{fig:SNi12TEM} showed highly crystalline particles spanning tens to hundreds of nanometres. The EDX analysis of particles in~\autoref{fig:A12EDX} confirmed the presence of oxygen and nickel in spherical and cubic particles.

\begin{figure}[htbp]
    \centering
    \includegraphics[width=0.8\linewidth]{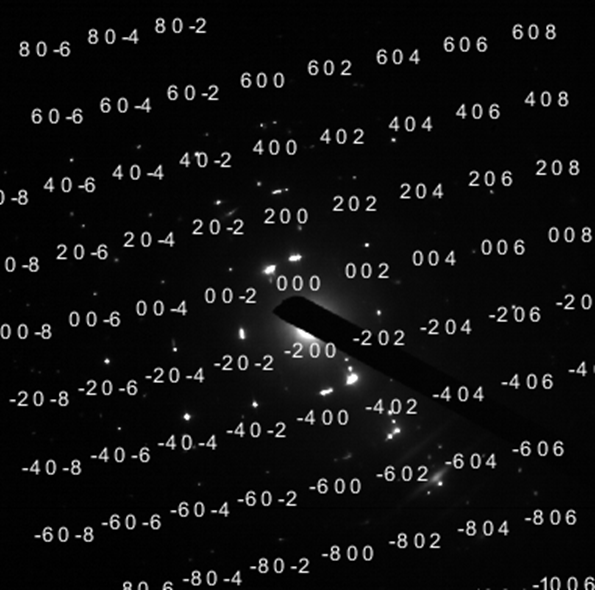}
    \caption{Electron diffraction of a particle created in the detonation of NHN matched to NiO ICSD \#9866.}
    \label{fig:A12ED}
\end{figure}

The electron diffraction pattern in~\autoref{fig:A12ED} is consistent with NiO.

\subsubsection{Nickel tris(ethylenediamine) perchlorate}

\begin{figure}[htbp]
    \centering
    \begin{minipage}[t]{0.48\textwidth}
        \centering
        \includegraphics[width=\linewidth]{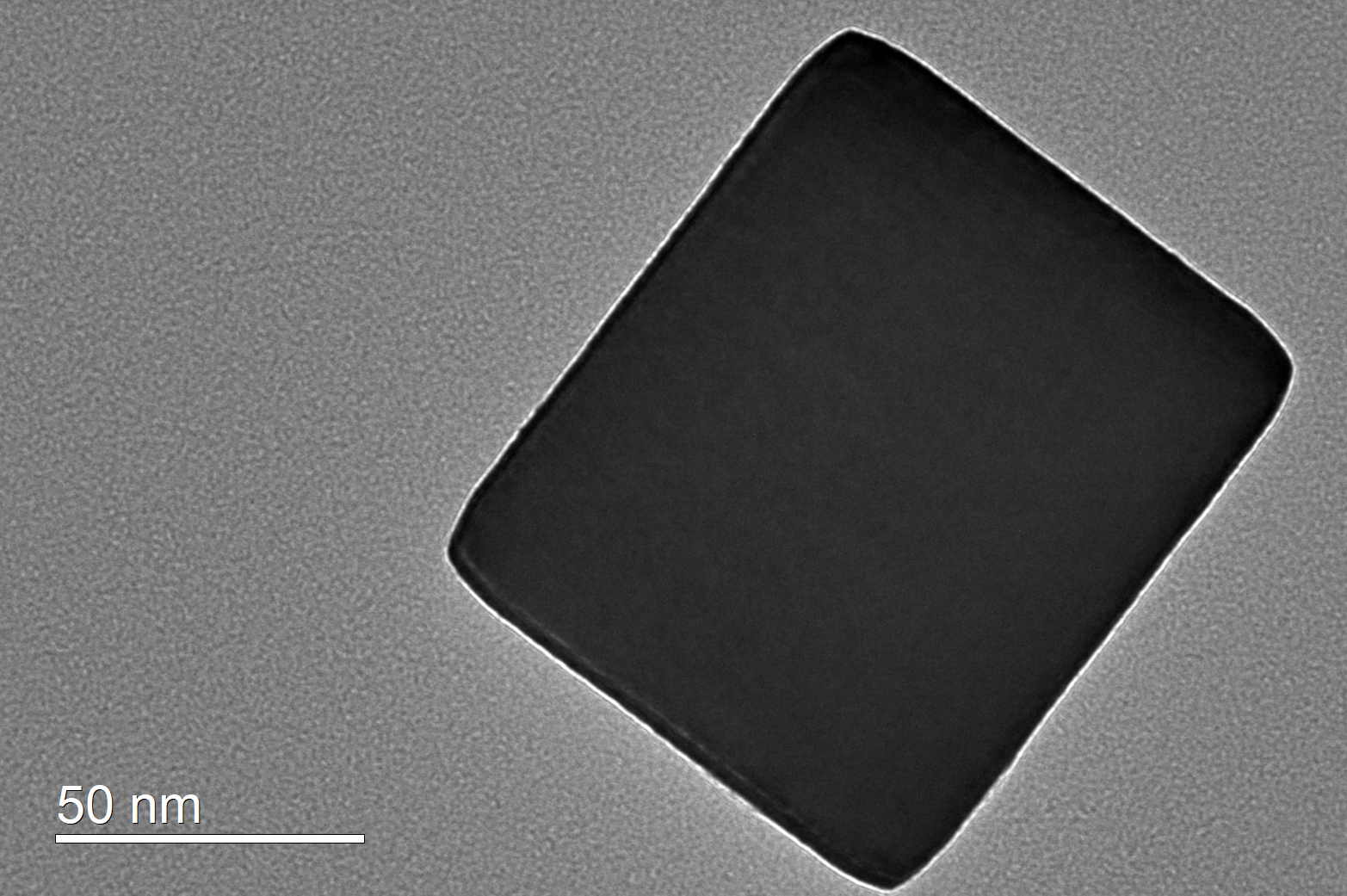}\\
        \textbf{(a)}
    \end{minipage}
    \hfill
    \begin{minipage}[t]{0.48\textwidth}
        \centering
        \includegraphics[width=\linewidth]{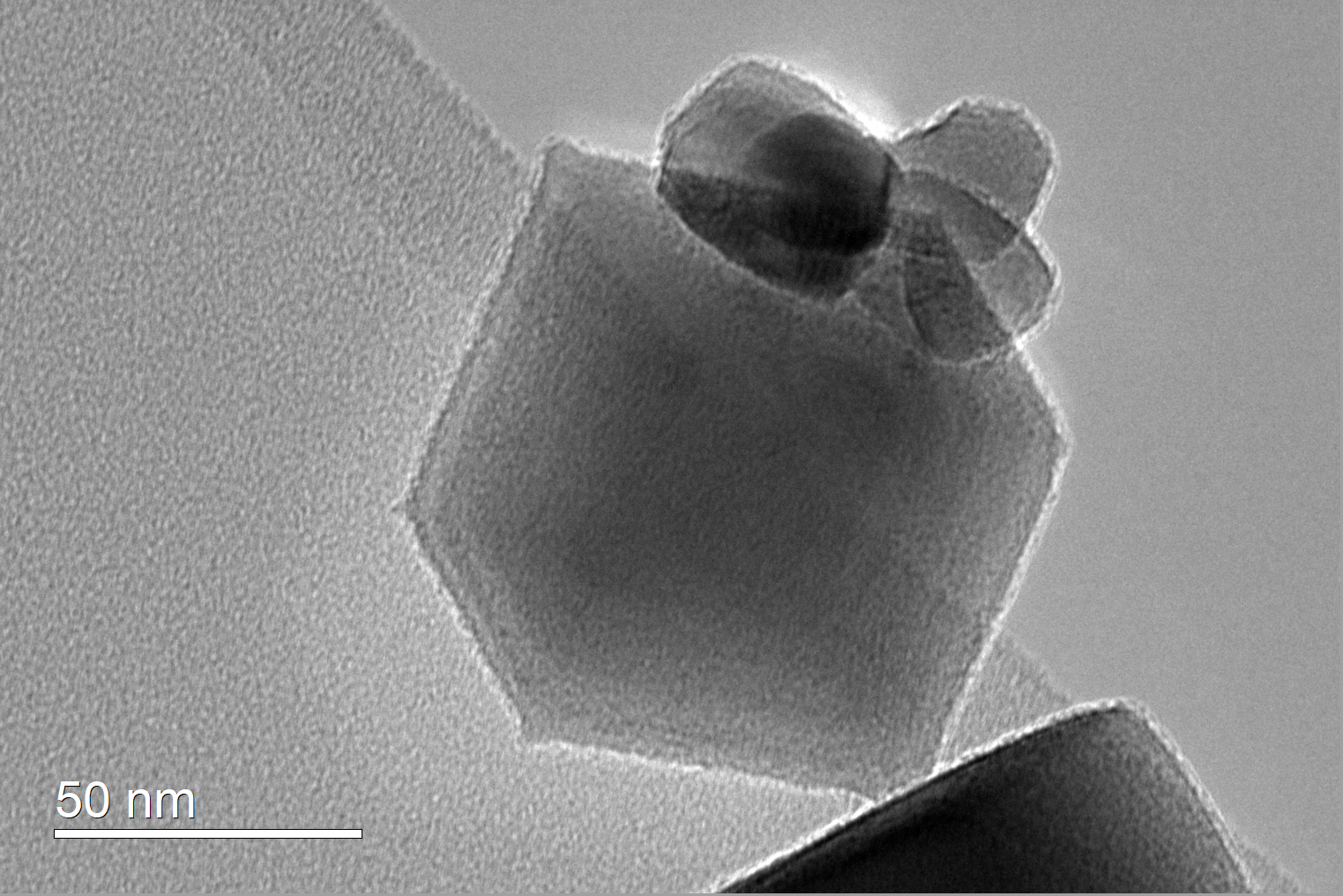}\\
        \textbf{(b)}
    \end{minipage}
    \caption{Representative particles produced by detonation of NiEP.}
    \label{fig:A14TEM}
\end{figure}

\begin{figure}[htbp]
    \centering
    \begin{minipage}[t]{0.2\textwidth}
        \centering
        \includegraphics[width=\linewidth]{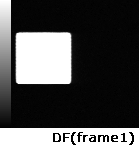}\\
        \textbf{(a)}
    \end{minipage}
    \hspace{0.04\textwidth}
    \begin{minipage}[t]{0.2\textwidth}
        \centering
        \includegraphics[width=\linewidth]{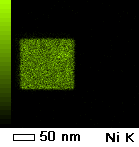}\\
        \textbf{(b)}
    \end{minipage}
    \hspace{0.04\textwidth}
    \begin{minipage}[t]{0.2\textwidth}
        \centering
        \includegraphics[width=\linewidth]{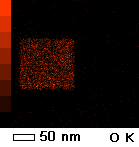}\\
        \textbf{(c)}
    \end{minipage}
    \caption{EDX map comparing the (a) dark-field image, (b) nickel content and (c) oxygen content of particles created in the detonation of NiEP.}
    \label{fig:A14EDX}
\end{figure}

TEM micrographs of particles produced by detonation of NiEP in~\autoref{fig:A14TEM} showed highly crystalline particles spanning tens to hundreds of nanometres. The EDX analysis of particles in~\autoref{fig:A14EDX} confirmed the presence of oxygen and nickel in the particles.

\begin{table}[htbp]
    \caption{Lattice fringe \textit{d}-spacing analysis of the A14 sample.}
    \label{tab:A14Lattice}
    \centering
    \begin{tabular}{c|cc}
        Measured & \multicolumn{2}{c}{NiO ICSD \#9866}\\
        \hline
        \textit{d}-spacing/\AA & \textit{d}-spacing/\AA & lattice plane\\
        \hline
        2.08 & 2.09 & ($200$)\\
    \end{tabular}
\end{table}

The \textit{d}-spacing measured is consistent with that of NiO as shown in~\autoref{tab:A14Lattice}.

\subsubsection{Nickel tris(carbohydrazide) perchlorate}

\begin{figure}[htbp]
    \centering
    \begin{minipage}[t]{0.48\textwidth}
        \centering
        \includegraphics[width=\linewidth]{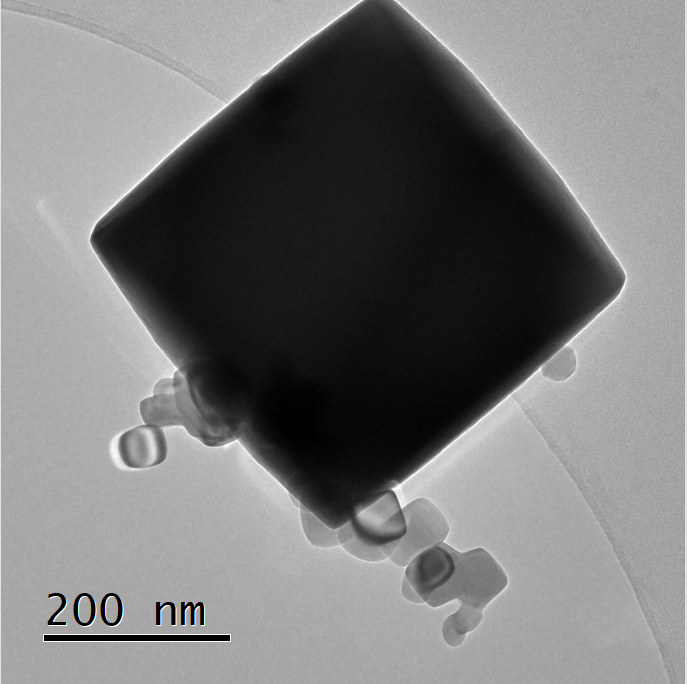}\\
        \textbf{(a)}
    \end{minipage}
    \hfill
    \begin{minipage}[t]{0.48\textwidth}
        \centering
        \includegraphics[width=\linewidth]{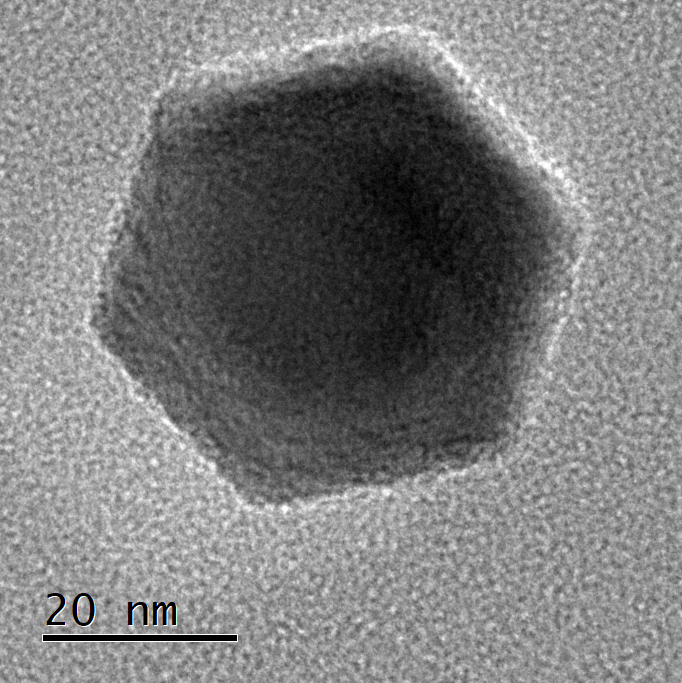}\\
        \textbf{(b)}
    \end{minipage}
    \caption{TEM micrographs showing particles formed in the detonation of NiCP.}
    \label{fig:A15TEM}
\end{figure}

\begin{figure}[htbp]
    \centering
    \begin{minipage}[t]{0.2\textwidth}
        \centering
        \includegraphics[width=\linewidth]{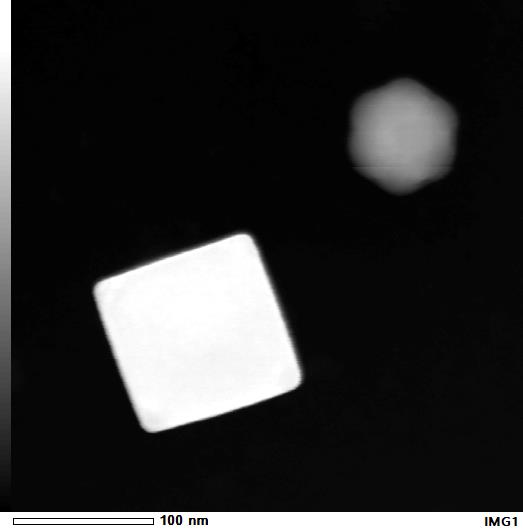}\\
        \textbf{(a)}
    \end{minipage}
    \hspace{0.04\textwidth}
    \begin{minipage}[t]{0.2\textwidth}
        \centering
        \includegraphics[width=\linewidth]{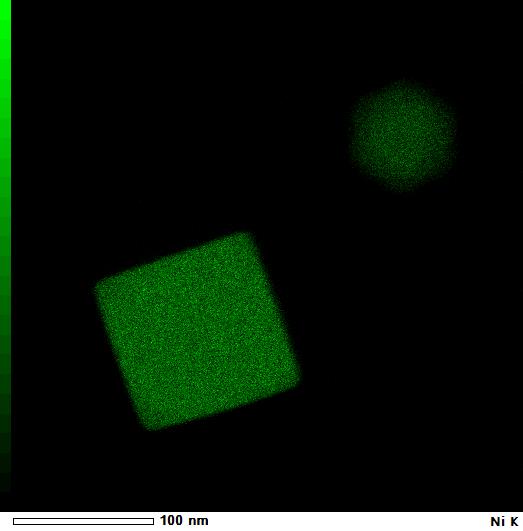}\\
        \textbf{(b)}
    \end{minipage}
    \hspace{0.04\textwidth}
    \begin{minipage}[t]{0.2\textwidth}
        \centering
        \includegraphics[width=\linewidth]{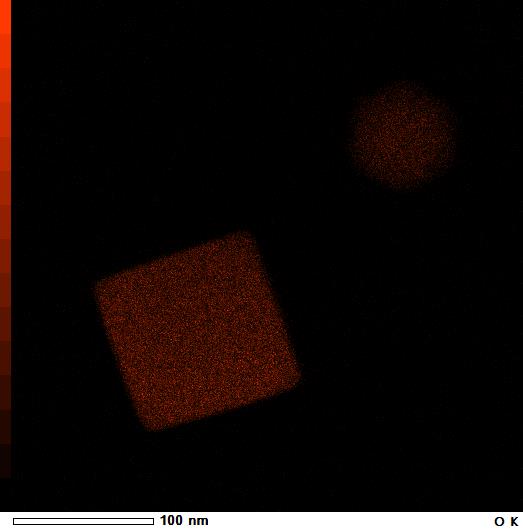}\\
        \textbf{(c)}
    \end{minipage}
    \caption{EDX map comparing the (a) dark-field image, (b) nickel content and (c) oxygen content of particles formed in the detonation of NiCP.}
    \label{fig:A15EDX}
\end{figure}

TEM micrographs of particles produced by detonation of NiCP in~\autoref{fig:A15TEM} showed examples of crystalline particles. The EDX analysis in~\autoref{fig:A15EDX} confirmed the presence of nickel and oxygen in both cubic and hexagonal particles.

\begin{figure}[htbp]
    \centering
    \includegraphics[width=.80\linewidth]{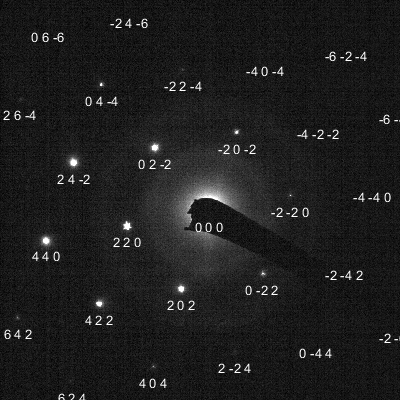}
    \caption{Indexed electron diffraction pattern from a hexagonal-appearing particle. The pattern is consistent with rock-salt NiO viewed along a cubic zone axis.}
    \label{fig:SNiED}
\end{figure}

The electron diffraction pattern in~\autoref{fig:SNiED} shows that the hexagonal-appearing particle is consistent with cubic NiO viewed along the $[\bar{1}1\bar{1}]$ direction, corresponding to a body diagonal between opposite vertices of the cubic structure.

The three chemically distinct nickel-containing energetic precursors produced Ni--O particles. The indexed diffraction pattern in~\autoref{fig:SNiED} demonstrates that a hexagonal-looking projection can arise from cubic NiO, supporting NiO as the common product across the nickel precursor set.
\subsection{Silver-containing energetic precursors and metastable Ag}

The silver-containing systems presented here expand on the particles reported by Uszko \textit{et al.} in 2025~\cite{Uszko2025Silver}. That study focused primarily on metallic nanoparticles produced by the BANG process, while a number of metal-oxide-containing products observed across the wider precursor library were not discussed in detail. The results presented here extend the scope of that work by demonstrating that BANG can also produce a range of silver oxide-containing nanoparticles, alongside metallic and metastable silver phases. The observed products additionally illustrate the influence of both precursor chemistry and detonation atmosphere on the phases retained after detonation.

\subsubsection{Silver azide}
\begin{figure}[htbp]
    \centering
    \begin{minipage}[t]{0.48\textwidth}
        \centering
        \includegraphics[width=\linewidth]{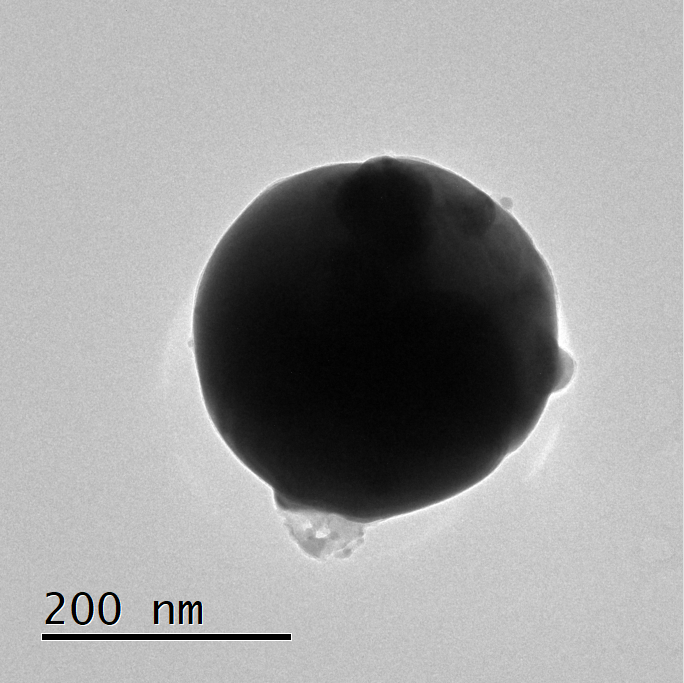}\\
        \textbf{(a)}
    \end{minipage}
    \hfill
    \begin{minipage}[t]{0.48\textwidth}
        \centering
        \includegraphics[width=\linewidth]{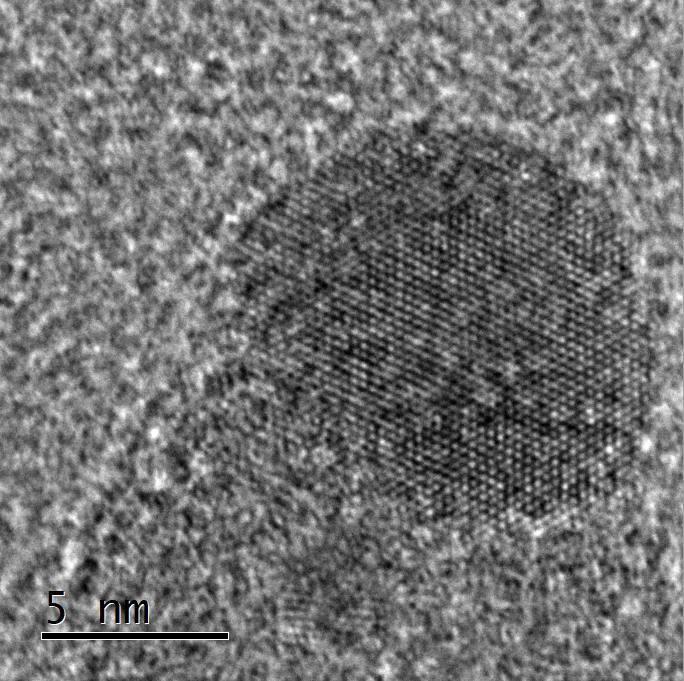}\\
        \textbf{(b)}
    \end{minipage}
    \caption{EM and HRTEM images of silver-containing particles produced by detonation of silver azide in air.}
    \label{fig:SAgA17TEM}
\end{figure}

Detonation of silver azide in air produced aggregated crystalline nanoparticles, as shown by the TEM and HRTEM images in~\autoref{fig:SAgA17TEM}. In contrast to the nitrogen-atmosphere experiment, local lattice-fringe measurements were consistent with an oxidised silver phase.

The measured spacings shown in~\autoref{tab:SAgA17Lattice} agree closely with reflections from Ag$_3$O$_4$, supporting the formation of silver oxide-containing nanoparticles during detonation in air.

\begin{table}[htbp]
    \caption{Local lattice-fringe comparison for particles produced by detonation of silver azide in air.}
    \label{tab:SAgA17Lattice}
    \centering
    \begin{tabular}{c|cc}
        Measured & \multicolumn{2}{c}{Ag$_3$O$_4$, ICSD \#202218}\\
        \hline
        \textit{d}-spacing/\AA & reference \textit{d}-spacing/\AA & lattice plane\\
        \hline
        2.35 & 2.35 & (022)\\
        2.78 & 2.75 & (120)\\
        2.44 & 2.41 & (11$\bar{2}$)\\
    \end{tabular}
\end{table}

\subsubsection{Silver 5,5'-azotetrazolate}
\begin{figure}[htbp]
    \centering
    \begin{minipage}[t]{0.48\textwidth}
        \centering
        \includegraphics[width=\linewidth]{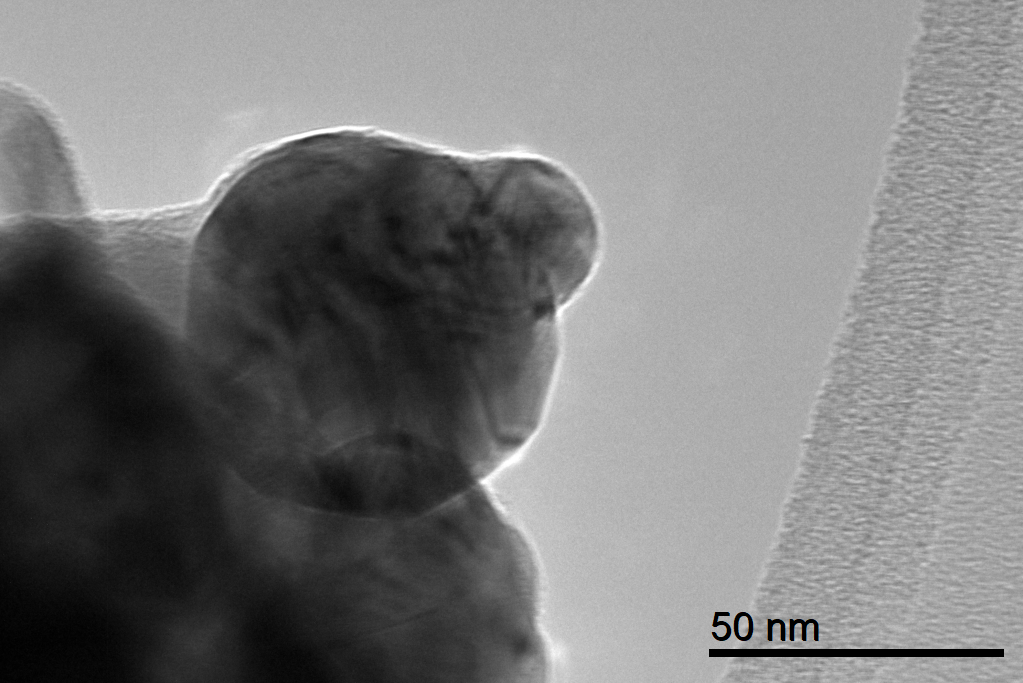}\\
        \textbf{(a)}
    \end{minipage}
    \hfill
    \begin{minipage}[t]{0.48\textwidth}
        \centering
        \includegraphics[width=\linewidth]{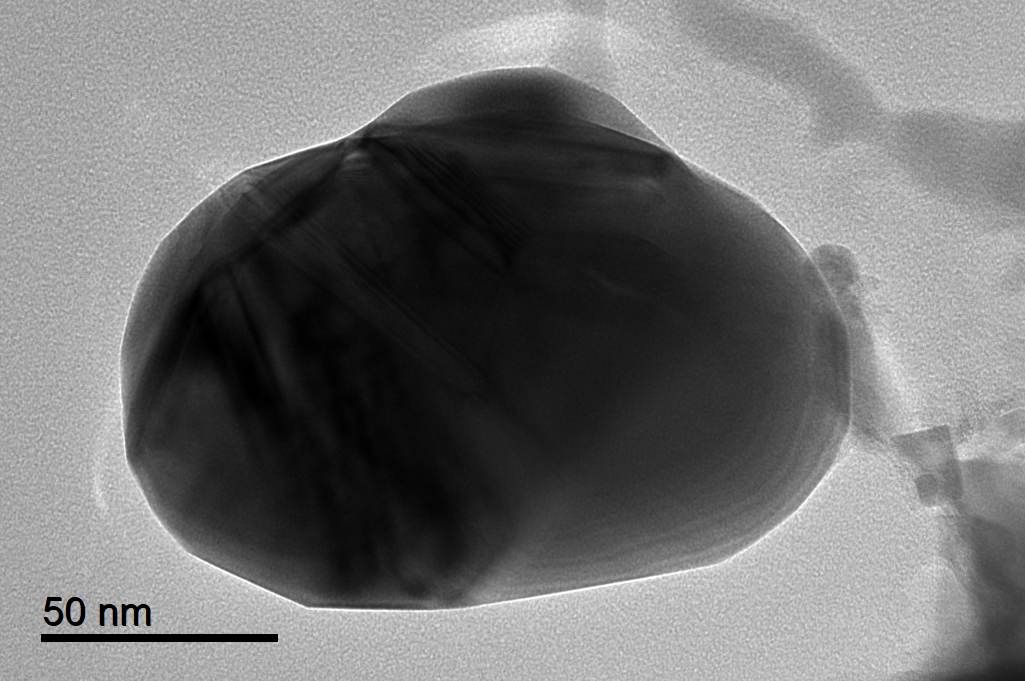}\\
        \textbf{(b)}
    \end{minipage}
    \caption{TEM images of silver-containing particles produced by detonation of silver 5-azotetrazole.}
    \label{fig:SAgC2TEM}
\end{figure}

Detonation of silver 5,5'-azotetrazolate similarly resulted in the formation of crystalline particles, representative examples of which are shown in~\autoref{fig:SAgC2TEM}.

Local lattice-fringe measurements from these particles were consistent with AgO, with the observed spacings corresponding closely to the reference values listed in~\autoref{tab:SAgC2Lattice}.

\begin{table}[htbp]
    \caption{Local lattice-fringe comparison for particles produced by detonation of silver 5-azotetrazole.}
    \label{tab:SAgC2Lattice}
    \centering
    \begin{tabular}{c|cc}
         & \multicolumn{2}{c}{AgO, ICSD \#202543}\\
        Measured/\AA & reference/\AA & plane \\
        \hline
        2.22 & 2.20 & (21$\bar{1}$)\\
        2.79 & 2.79 & (200)\\
        1.13 & 1.13 & (13$\bar{1}$)\\
    \end{tabular}
\end{table}

\subsubsection{Silver 5-nitrotetrazolate}
\begin{figure}[htbp]
    \centering
    \begin{minipage}[t]{0.48\textwidth}
        \centering
        \includegraphics[width=\linewidth]{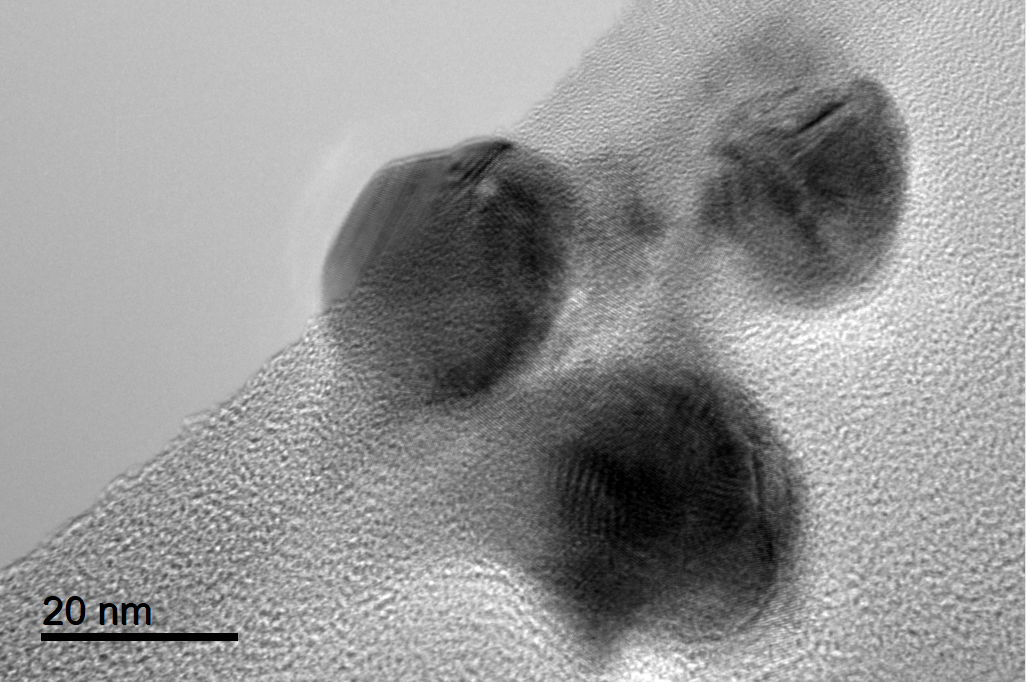}\\
        \textbf{(a)}
    \end{minipage}
    \hfill
    \begin{minipage}[t]{0.48\textwidth}
        \centering
        \includegraphics[width=\linewidth]{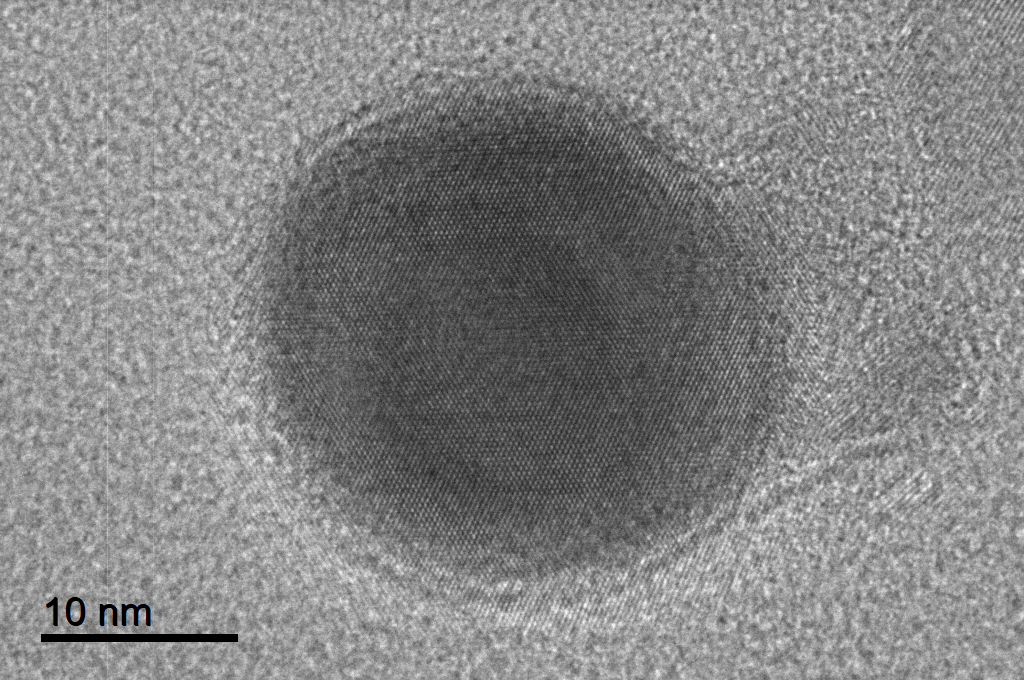}\\
        \textbf{(b)}
    \end{minipage}
    \caption{TEM images of silver-containing particles produced by detonation of silver 5-nitrotetrazolate.}
    \label{fig:SAgC5TEM}
\end{figure}

TEM micrographs in~\autoref{fig:SAgC5TEM} show particles prepared in the detonation of silver 5-nitrotetrazolate.

As shown in~\autoref{tab:SAgC5Lattice}, the measured spacings are close to reflections from both AgO and Ag$_3$O$_4$. These local measurements therefore support the presence of oxidised silver phases, although they do not provide a unique bulk-phase assignment.

\begin{table}[htbp]
    \caption{Local lattice-fringe comparison for particles produced by detonation of silver 5-nitrotetrazole.}
    \label{tab:SAgC5Lattice}
    \centering
    \setlength{\tabcolsep}{3pt}
    \small
    \begin{tabular}{c|cc|cc}
        & \multicolumn{2}{c|}{\shortstack{AgO\\ICSD \#202543}}
        & \multicolumn{2}{c}{\shortstack{Ag$_3$O$_4$\\ICSD \#202218}}\\
        Measured/\AA
        & reference/\AA & plane
        & reference/\AA & plane\\
        \hline
        2.36 & 2.42  & (111)                 & 2.35  & (022)\\
        2.80 & 2.79  & (200)                 & 2.75  & (120)\\
        1.58 & 1.595 & $(31\bar{2})$     & 1.578 & $(22\bar{2})$\\
    \end{tabular}
\end{table}

Together, these observations expand the range of products described in the 2025 study by Uszko \textit{et al.} beyond the predominantly metallic nanoparticles discussed there. The additional silver systems demonstrate that BANG can also access oxide-containing nanoparticles, while the contrast between silver azide detonated in air and under nitrogen illustrates the influence of the detonation atmosphere on the phases retained in the resulting material.
\subsubsection{Silver azide detonated under nitrogen}
\begin{figure}[!t]
    \centering

    \subfloat[\textbf{(a)}\label{fig:HCPAg_TEM}]{%
        \begin{minipage}[b]{0.4\textwidth}
        \centering
        \includegraphics[width=\linewidth]
        {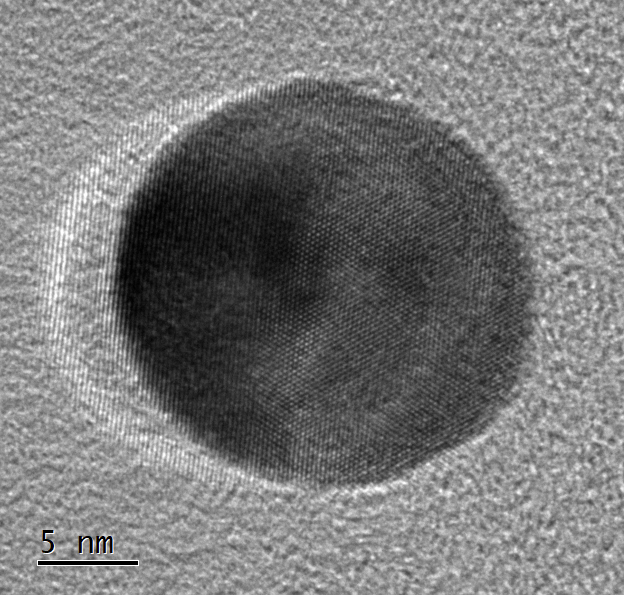}
        \end{minipage}%
    }
    \hfill
    \subfloat[\textbf{(b)}\label{fig:HCPAg_FFT}]{%
        \begin{minipage}[b]{0.4\textwidth}
        \centering
        \includegraphics[width=\linewidth]
        {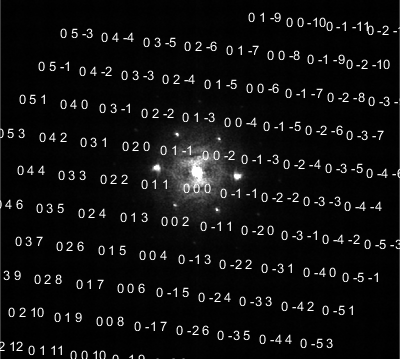}
        \end{minipage}%
    }

    \caption{Characterisation of silver nanoparticles produced by detonation
    of silver azide under a nitrogen atmosphere.
    (a) TEM micrograph of a representative nanoparticle.
    (b) FFT obtained from the particle, with selected reflections indexed
    to the corresponding lattice planes of HCP Ag.}
    \label{fig:HCP_silver}
\end{figure}

In EEW, the use of a non-oxidising atmosphere can suppress oxidation, enabling metallic nanoparticles to be produced from metals that would otherwise form oxides or mixed metal/oxide products.~\cite{Kotov_2003} Previous work by Uszko \textit{et al.} showed that upon detonation of silver-containing explosives in air, a mixture of face-centred cubic (FCC) silver and hexagonal close-packed (HCP) silver was produced. Further characterisation of particles produced in air revealed local lattice spacings consistent with the presence of silver oxides. Upon detonation of silver azide under nitrogen, particles containing the rare HCP polytype of silver were observed,~\autoref{fig:HCP_silver}, alongside the more typical FCC; however, no silver oxide particles were detected. The indexed fast Fourier transform (FFT) is consistent with HCP Ag, while the corresponding lattice-spacing analysis is provided in~\autoref{tab:SAgHCPLattice}. EEW has also been reported to produce metastable phases of iron and tungsten.~\cite{Kotov_2003}
The locally measured lattice spacings are consistent with reflections expected for HCP Ag, as summarised in~\autoref{tab:SAgHCPLattice}. The observation of HCP Ag under nitrogen is consistent with rapid formation and cooling during detonation allowing a metastable silver polytype to be retained.

\begin{table}[htbp]
    \caption{Local lattice-fringe measurements for silver nanoparticles produced by detonation of silver azide under nitrogen.}
    \label{tab:SAgHCPLattice}
    \centering
    \begin{tabular}{c|cc}
        Measured & \multicolumn{2}{c}{HCP Ag, ICSD \#56269}\\
        \hline
        \textit{d}-spacing/\AA & reference \textit{d}-spacing/\AA & lattice plane\\
        \hline
        2.32 & 2.40 & $(0002)$\\
        2.22 & 2.24 & $(10\bar{1}1)$\\
        1.33 & 1.35 & $(10\bar{1}3)$\\
        2.18 & 2.24 & $(10\bar{1}1)$\\
    \end{tabular}
\end{table}

\subsection{Cadmium-containing energetic precursor}

\begin{figure}[htbp]
    \centering
    \begin{minipage}[t]{0.48\textwidth}
        \centering
        \includegraphics[width=\linewidth]{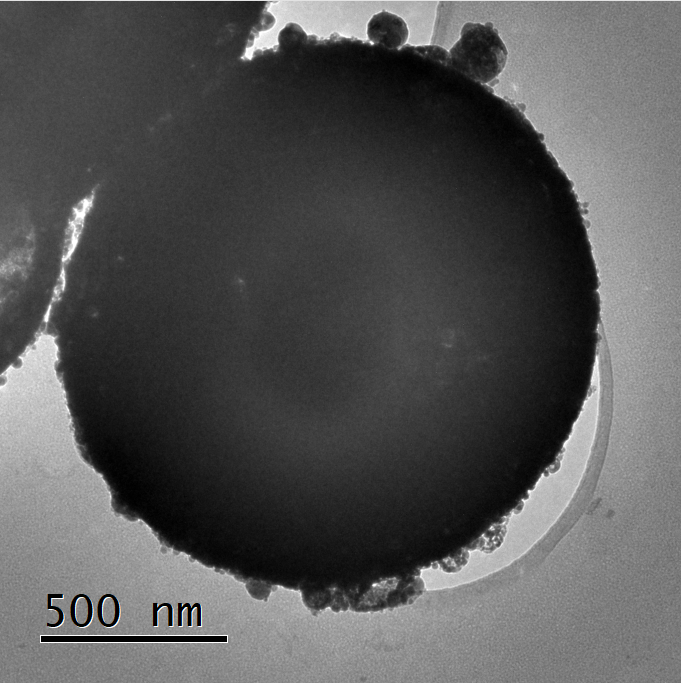}\\
        \textbf{(a)}
    \end{minipage}
    \hfill
    \begin{minipage}[t]{0.48\textwidth}
        \centering
        \includegraphics[width=\linewidth]{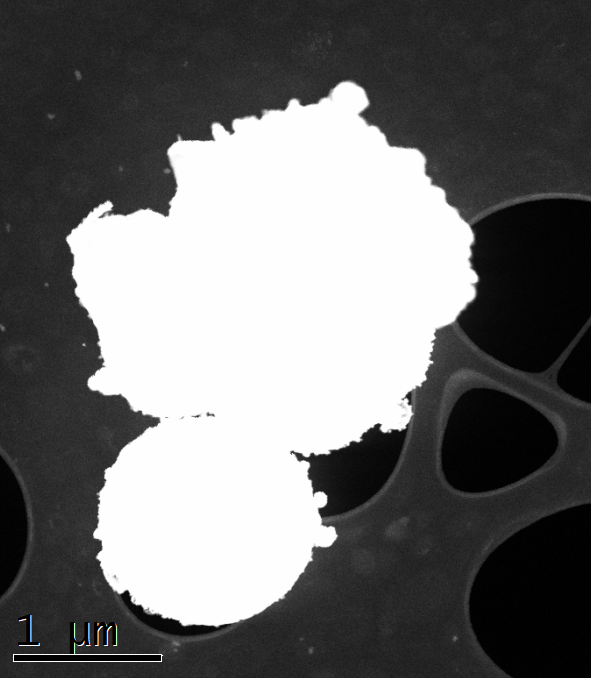}\\
        \textbf{(b)}
    \end{minipage}
    \caption{Bright-field (a) and dark-field (b) TEM images of particles produced by detonation of cadmium 5,5'-azotetrazolate complex.}
    \label{fig:SCdTEM}
\end{figure}

\begin{figure}[htbp]
    \centering
    \begin{minipage}[t]{0.20\textwidth}
        \centering
        \includegraphics[width=\linewidth]{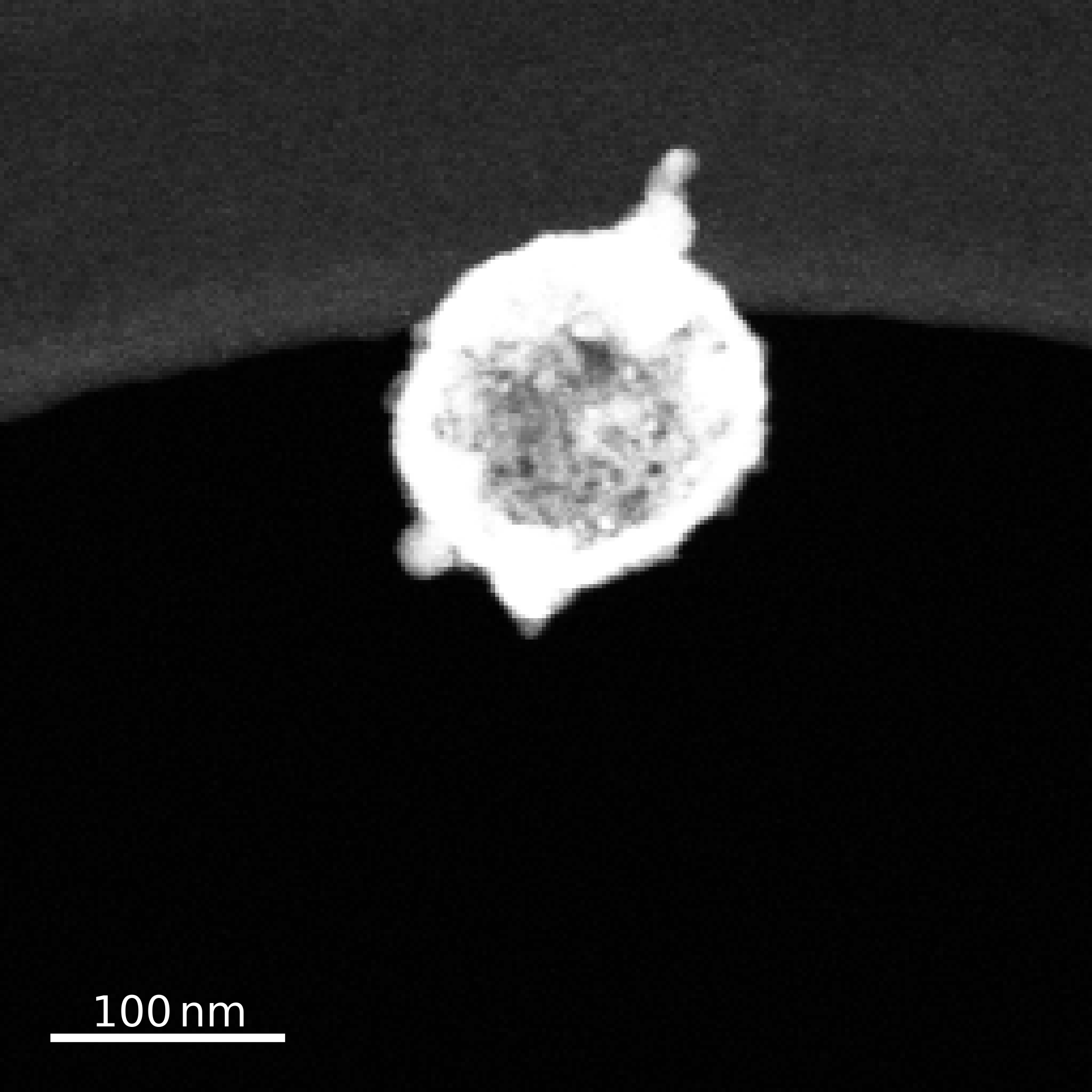}\\
        \textbf{(a)}
    \end{minipage}
    \begin{minipage}[t]{0.20\textwidth}
        \centering
        \includegraphics[width=\linewidth]{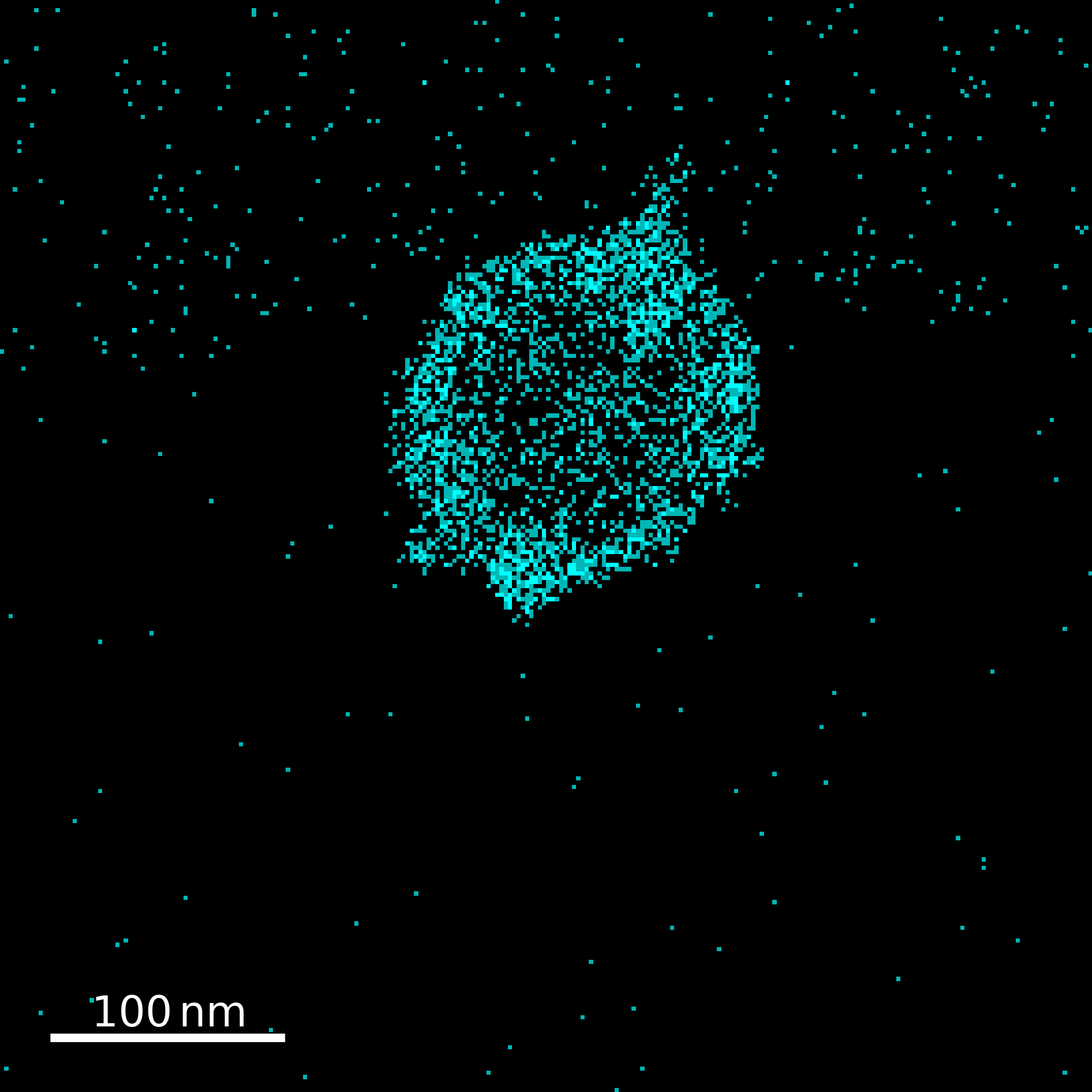}\\
        \textbf{(b)}
    \end{minipage}
    \hfill
    \begin{minipage}[t]{0.20\textwidth}
        \centering
        \includegraphics[width=\linewidth]{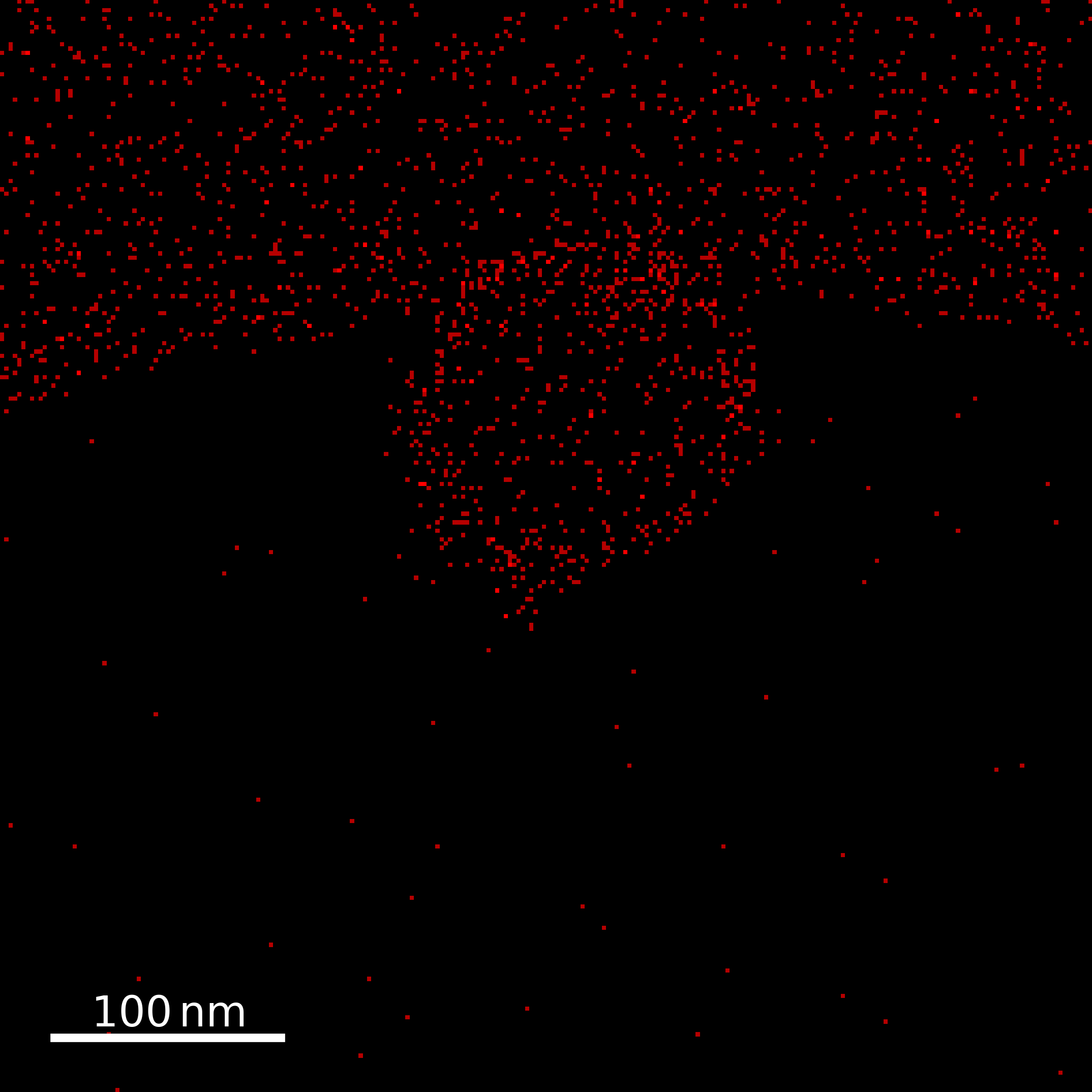}\\
        \textbf{(c)}
    \end{minipage}
    \caption{Electron micrograph (a) and corresponding EDX maps for
    (b) Cd and (c) O.}
    \label{fig:SCdEDX}
\end{figure}

TEM micrographs of particles produced by detonation of cadmium 5,5'-azotetrazolate complex in~\autoref{fig:SCdTEM} showed highly crystalline particles spanning tens to hundreds of nanometres. The EDX analysis of particles in~\autoref{fig:SCdEDX} confirmed the presence of cadmium as well as surface-level oxygen.

\begin{table}[htbp]
    \caption{Local lattice-fringe measurements for the cadmium-containing sample.}
    \label{tab:SCdLattice}
    \centering
    \begin{tabular}{c|cc}
        Measured & \multicolumn{2}{c}{Cd, ICSD \#619639}\\
        \hline
        \textit{d}-spacing/\AA &
        reference \textit{d}-spacing/\AA &
        lattice plane\\
        \hline
        2.81 & 2.81 & (0002)\\
        2.33 & 2.34 & (10$\bar{1}$1)\\
    \end{tabular}
\end{table}

The measured lattice spacings in~\autoref{tab:SCdLattice} of 2.81 and 2.33~\AA\ are consistent with metallic Cd, corresponding to the $(0002)$ and $(10\bar{1}1)$ reflections, respectively. The latter spacing is also close to the CdO $(200)$ reflection and suggests that a thin layer of oxide formed on the surface.
\subsection{Palladium-containing energetic precursor and direct deposition}
\begin{figure}[htbp]
    \centering
    \begin{minipage}[t]{0.20\textwidth}
        \centering
        \includegraphics[width=\linewidth]{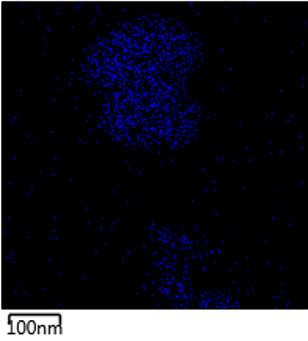}\\
        \textbf{(a)}
    \end{minipage}
    \hspace{0.04\textwidth}
    \begin{minipage}[t]{0.20\textwidth}
        \centering
        \includegraphics[width=\linewidth]{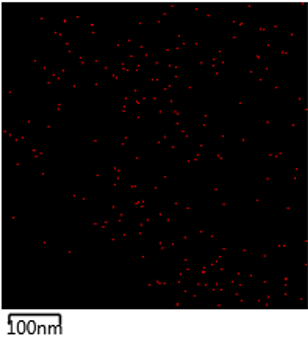}\\
        \textbf{(b)}
    \end{minipage}
    \caption{EDX maps showing (a) Pd and (b) O distributions in particles produced by detonation of palladium 5,5'-azotetrazolate.}
    \label{fig:SPdEDX}
\end{figure}

\begin{figure}[htbp]
    \centering

    \subfloat[\textbf{(a)}\label{fig:Pd_TEM}]{%
        \begin{minipage}[b]{0.45\textwidth}
        \centering
        \includegraphics[width=\linewidth]{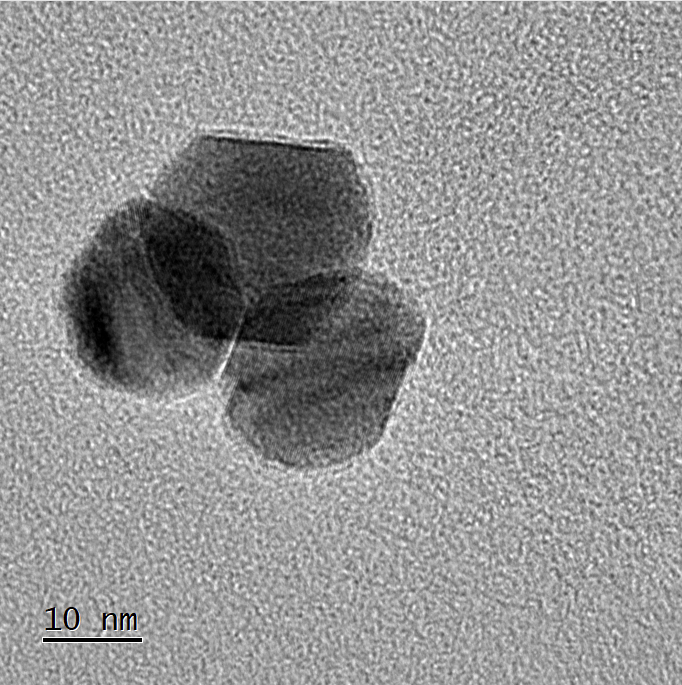}
        \end{minipage}%
    }
    \hfill
    \subfloat[\textbf{(b)}\label{fig:Pd_carbon}]{%
        \begin{minipage}[b]{0.45\textwidth}
        \centering
        \includegraphics[width=\linewidth]{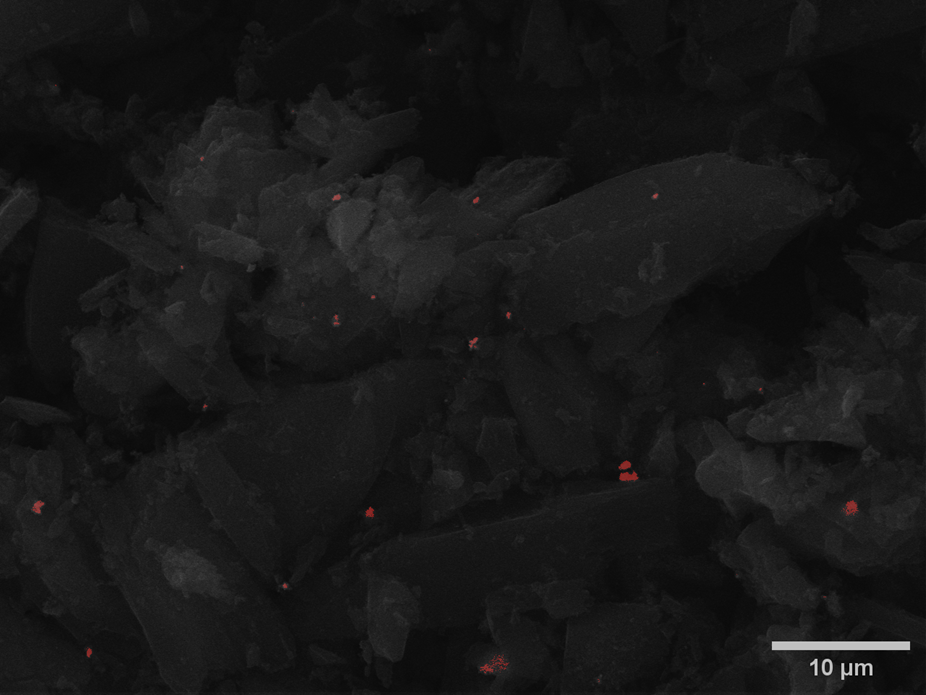}
        \end{minipage}%
    }

    \caption{BANG-derived palladium nanoparticles and direct deposition onto a support.
    (a) TEM image of crystalline particles produced by detonation of palladium
    5,5'-azotetrazolate. (b) SEM image of activated carbon following direct
    deposition of Pd-containing nanoparticles during the BANG process with palladium marked in red.}
    \label{fig:Pd_deposition}
\end{figure}

TEM micrographs of particles produced by detonation of palladium 5,5'-azotetrazolate in~\autoref{fig:Pd_deposition}a showed examples of crystalline particles. EDX mapping in~\autoref{fig:SPdEDX} confirmed the presence of palladium and showed a lack of oxygen in the analysed particles.

One of the earliest reported uses of fulminating gold was gilding.~\cite{Glauber1658,Wentrup2019} Detonation propelled created particles onto the substrate. Similar direct-deposition approaches have been reported for EEW and for BANG in the production of silver films.~\cite{Mattox_1998,Uszko2025Silver} To test the possibility of utilisation of BANG as a deposition method a sample of palladium 5,5'-azotetrazolate was prepared. Its detonation produced highly crystalline, faceted Pd-containing nanoparticles, as shown in~\autoref{fig:Pd_deposition}a. The BANG process was also used to deposit Pd-containing nanoparticles directly onto activated carbon,~\autoref{fig:Pd_deposition}b.

Direct deposition onto activated carbon demonstrates that BANG can produce supported nanoparticles without an intermediate particle-isolation or deposition step, providing a potential route towards supported functional materials.
Palladium particles were deposited onto an activated carbon substrate as described in the~\autoref{sec:depo}. An SEM image of the deposited particles is shown in~\autoref{fig:Pd_deposition}.

The SEM-based analysis indicated an approximately 0.13\% Pd signal in the analysed supported material. This value should be described as a local/semiquantitative SEM-based estimate rather than a rigorous bulk metal loading.

\section{Experimental Methods}

\subsection{Material Preparation}
Where sufficient material was available, the prepared compounds were characterised by FT-IR spectroscopy, differential thermal analysis (DTA) and elemental analysis. Owing to the energetic nature of the materials and the limited quantities available, these analyses could not be performed for all compounds.
\subsubsection{Lead azide}
Lead azide was provided by the Austin Detonator Vsetín Company, Czech Republic. The sample was not analysed.

\subsubsection{Lead picrate}
Lead picrate was provided by the Reichel \& Partner Company, Austria. The sample was not analysed.
\subsubsection{Lead styphnate}
Lead styphnate was provided by the Sellier \& Bellot Company, Czech Republic. The sample was not analysed.

\subsubsection{Basic bismuth 5,5'-azotetrazolate}
Basic bismuth 5,5'-azotetrazolate was provided by the Sellier \& Bellot Company, Czech Republic.

\subsubsection{Thallium azide}
Thallium azide was prepared by precipitation from aqueous solutions of thallous sulfate (504 mg, 1 mmol) in distilled water (20 ml) and sodium azide (130 mg, 2 mmol) in distilled water (2.6 ml). Upon mixing the solutions under continuous stirring, an immediate white precipitate formed. The suspension was stirred for 10 min to ensure complete precipitation, after which the solid was isolated by filtration, washed with distilled water, and dried. Thallium azide was obtained as a white crystalline powder, which gradually darkened during drying as a result of progressive surface photolysis.

Analysis: FT-IR (cm$^{-1}$): 3259m, 3195w, 3176w, 3158w, 1916m, 1417w, 628m, 622m. Elemental analysis: Calc. for N$_3$Tl (\%): N 17.06, Tl 82.94. Found: H 0.07, N 16.53. DTA (5$^{\circ}$C min$^{-1}$): 336$^{\circ}$C (melting).

\subsubsection{Manganese tris(carbohydrazide) perchlorate (MnCP)}
Manganese tris(carbohydrazide) perchlorate (MnCP) was prepared according to a modified version of the procedure published by Talawar \textit{et al.}, in which nickel perchlorate was replaced with manganese(II) perchlorate hexahydrate.~\cite{Talawar_2004}

Analysis: FT-IR (cm$^{-1}$): 3355m, 3323m, 1644s, 1597w, 1540m, 1446w, 1425m, 1207w, 1191w, 1145w, 1059s, 961w, 933w, 758m, 692w, 614s. Elemental analysis: Calc. for C$_3$H$_{18}$N$_{12}$O$_{11}$Cl$_2$Mn (\%): C 6.88, H 3.46, N 32.07. Found: C 7.15, H 3.32, N 32.43. DTA (5~$^\circ$C\,min$^{-1}$): 230$^\circ$C (start of decomposition), 322$^\circ$C (maximum). Metal analysis: calculated Mn 10.48\%; found Mn 10.30\%.

\subsubsection{Iron tris(carbohydrazide) perchlorate (FeCP)}
Iron tris(carbohydrazide) perchlorate (FeCP) was prepared according to a modified version of the procedure published by Talawar \textit{et al.}, in which nickel perchlorate was replaced with iron(II) perchlorate hexahydrate.~\cite{Talawar_2004}

Analysis: FT-IR (cm$^{-1}$): 3350m, 3302m, 1643s, 1542m, 1444w, 1423m, 1205m, 1147w, 1060s, 963w, 907w, 756m, 694w. Elemental analysis: Calc. for C$_3$H$_{18}$N$_{12}$O$_{11}$Cl$_2$Fe (\%): C 6.86, H 3.46, N 32.02. Found: C 7.50, H 3.42, N 32.88. DTA (5~$^\circ$C\,min$^{-1}$): 150$^\circ$C (start of decomposition), 160$^\circ$C (maximum). Metal analysis: calculated Fe 10.64\%; found Fe 11.93\%.
\subsubsection{Tetraamminecopper perchlorate (TACP)}
Tetraamminecopper perchlorate (TACP) was prepared according to the procedure of Vodochodský \textit{et al.},~\cite{Vodochodsk__2020}  with the synthesis performed on a 20-fold reduced scale. Copper perchlorate hexahydrate (5~g, 13.5~mmol) was dissolved in aqueous ammonia (12.5~ml, 167~mmol) while being stirred. The temperature was raised to 35~$^{\circ}$C, the solution was allowed to cool slowly to ambient temperature, and it was then cooled to 0~$^{\circ}$C using an ice bath. The crystallised dark blue product was filtered and left to dry freely at room temperature.

Analysis: FT-IR (cm$^{-1}$): 3354m, 3285w, 3199vw, 1615w, 1426w, 1253m, 1043s, 931m, 685m, 615s, 459w, 433w. Elemental analysis: Calc. for H$_{12}$N$_4$O$_8$CuCl$_2$ (\%): H 3.66, N 16.95. Found: C 0.1, H 3.58, N 16.22. DTA (5$^{\circ}$C min$^{-1}$): 233$^{\circ}$C (start of decomposition), 268$^{\circ}$C (explosion). Metal analysis: calculated Cu 19.22\%; found Cu 18.11\%.

\subsubsection{Cuprous acetylide}
Cuprous acetylide was prepared according to the procedure published by Špičák and Šimeček.~\cite{Spicak1957Chemie} Copper(II) chloride (10.0 g, 74.3 mmol) was dissolved in water (50 ml), after which five drops of 35\% hydrochloric acid and copper powder (10.0 g, 157 mmol) were added. The mixture was heated under reflux to reduce Cu(II) to Cu(I). Completion of the reduction was indicated by loss of colour and precipitation of white copper(I) chloride. Aqueous ammonia (25\%) was subsequently added until the copper(I) chloride had dissolved, and the remaining copper powder was removed by filtration. The resulting clear solution was then treated with a stream of  acetylene.~\cite{Maty__2012}.

Analysis: FT-IR (cm$^{-1}$): 3321m, 1456m, 1373s, 830w. Elemental analysis: Calc. for C$_2$Cu$_2$ (\%): C 15.89, Cu 84.11 Found: C 11.68, H 0.098. DTA (5$^{\circ}$C min$^{-1}$): 85$^{\circ}$C (start of decomposition), 249$^{\circ}$C (maximum).

\subsubsection{Zinc 5,5'-azotetrazolate dihydrate (ZnAzTZ$\cdot$2H$_2$O)}
A solution of zinc acetate dihydrate (1.72 g, 7.8 mmol) in water (60 ml) and a solution of sodium 5,5'-azotetrazolate pentahydrate (1.78 g, 5.9 mmol) in water (60 ml) were added simultaneously to water (60 ml) maintained at 70$^{\circ}$C under mechanical stirring. The zinc acetate solution was added slightly in excess throughout the addition. An immediate precipitate formed. After complete addition, the suspension was stirred for 15 min at 70$^{\circ}$C, cooled, and filtered. The product was washed with cold water and ethanol and dried to yield zinc 5,5'-azotetrazolate dihydrate.

Analysis: FT-IR (cm$^{-1}$): 3525m, 3461m, 3323m, 1612w, 1469w, 1433, 1415m, 1236w, 1205w, 1196w, 1165w, 964m, 777m, 769m, 746s, 589m, 578m. Elemental analysis: Calc. for C$_2$H$_4$N$_{10}$O$_2$Zn (\%): C 9.05, H 1.52, N 52.75, Zn 24.63. Found: C 8.70, H 1.85, N 46.30. DTA (5$^{\circ}$C min$^{-1}$): 186$^{\circ}$C (start of decomposition), 205$^{\circ}$C (maximum).

\subsubsection{Platinum-containing azide precursor}

A platinum-containing azide precursor was prepared from potassium tetrachloroplatinate(II) (30 mg, 0.072 mmol) and excess aqueous sodium azide. The recovered material consisted of a mixture of white and orange crystals. Platinum--azide coordination chemistry has previously been reported by Schmidtke and Garthoff.~\cite{Schmidtke1967Azide}

\subsubsection{Nickel tris(hydrazine) dinitrate (NHN)}
Nickel tris(hydrazine) dinitrate (NHN), Ni(N$_2$H$_4$)$_3$(NO$_3$)$_2$, was prepared according to the procedure reported by Shunguan \textit{et al.}~\cite{Shunguan_1997} Briefly, an aqueous nickel nitrate solution was reacted with hydrazine hydrate under controlled stirring and heating. Following completion of the reaction, the resulting suspension was cooled and the solid product was isolated by filtration, washed with water, treated with alcohol and dried. The published procedure, including reagent concentrations, quantities, addition conditions and isolation parameters, was followed as described in the original reference.
Analysis: FT-IR (cm$^{-1}$): 3302m, 3236m, 3179m, 1761w, 1625m, 1415w, 1389m, 1342s, 1310s, 1204m, 1181s, 1046w, 976m, 820m, 716w, 646m. Elemental analysis: Calc. for H$_{12}$N$_8$O$_6$Ni (\%): H 4.34, N 40.19. Found: H 4.07, N 39.21. DTA (5$^{\circ}$C min$^{-1}$): 215$^{\circ}$C (start of decomposition), 228$^{\circ}$C (maximum). Metal analysis: calculated Ni 21.05\%; found Ni 20.84\%.

\subsubsection{Nickel tris(ethylenediamine) perchlorate (NiEP)} 
Complex [Ni(en)$_3$](ClO$_4$)$_2$ was prepared according to a modified procedure by Lisensky.~\cite{Lisensky_2022} Nickel(II) perchlorate hexahydrate (3.66 g, 0.01 mol) was dissolved in 30 ml of ethanol at room temperature to afford a green solution. Subsequently, ethylenediamine (2.34 ml, 0.035 mol) was added dropwise under continuous stirring. During the addition, the colour of the reaction mixture gradually changed from dark green through blue to violet-pink, accompanied by the progressive precipitation of a pale violet solid. After complete addition of ethylenediamine, the reaction mixture was stirred for an additional 10 min and then cooled in an ice bath. The solid product was collected by vacuum filtration using a Büchner funnel, washed twice with small portions of cold ethanol, and allowed to dry in air in a fume hood.
Analysis: FT-IR (cm$^{-1}$): 3352m, 3302m, 2953w, 2896w, 1578m, 1464w, 1326w, 1281w, 1061s, 1011s, 969w, 931w, 879w, 653w, 619s. Elemental analysis: Calc. for C$_6$H$_{24}$N$_6$O$_8$Cl$_2$Ni (\%): C 16.46, H 5.52, N 19.19. Found: C 16.62, H 5.33, N 19.36. DTA (5$^{\circ}$C min$^{-1}$): 230$^{\circ}$C (start of decomposition), 283$^{\circ}$C (maximum). Metal analysis: calculated Ni 13.40\%; found Ni 12.06\%.

\subsubsection{Nickel tris(carbohydrazide) perchlorate (NiCP)}
Nickel tris(carbohydrazide) perchlorate (NiCP) was prepared according to the procedure published by Talawar \textit{et al.}.~\cite{Talawar_2004} A solution was prepared by dissolving 6.31 g (0.07 mol) of carbohydrazide in 35 ml of water and heating it to 58--60~$^{\circ}$C under mechanical stirring. A second solution containing 7.32 g (0.02 mol) of nickel perchlorate hexahydrate in 25--30 ml of DI water was added slowly over 20 min while maintaining the temperature at 58--60~$^{\circ}$C. The reaction mixture was stirred for a further 30 min at the same temperature, then cooled to room temperature and stirred for 1 h. The resulting light-blue NiCP crystals were isolated by vacuum filtration, washed with distilled water (4 $\times$ 50 ml) followed by 50 ml of propyl alcohol, and dried at 60~$^{\circ}$C.Metal analysis: calculated Ni 11.12\%; found Ni 10.52\%.

Analysis: FT-IR (cm$^{-1}$): 3343m, 3308m 1643s, 1600w, 1543m, 1444w, 1424m, 1211m, 1195w, 1150w, 1062s, 963w, 935w, 907w, 756m, 701w, 619s. Elemental analysis: Calc. for C$_3$H$_{18}$N$_{12}$O$_{11}$Cl$_2$Ni (\%): C 6.83, H 3.44, N 31.84. Found: C 6.85, H 3.33, N 32.33. DTA (5~$^\circ$C\,min$^{-1}$): 275$^\circ$C (start of decomposition), 306$^\circ$C (maximum).

\subsubsection{Silver azide}
Silver azide was prepared by a precipitation method. A 1 ml aqueous solution of sodium azide (50 mg, 0.76 mmol) was mixed with a solution of silver nitrate (23 mg, 0.13 mmol) in 0.5 ml of deionised water. This resulted in precipitation of silver azide, which was separated using vacuum filtration.~\cite{Matyas2013}
\subsubsection{Silver 5-nitrotetrazolate}
Sodium 5-nitrotetrazolate dihydrate (433 mg, 2.5 mmol) was dissolved in 30 ml of water in a beaker wrapped in aluminum foil to protect the reaction mixture from light. A solution of silver nitrate (510 mg, 3 mmol) in 10 mL of water was then added dropwise under continuous stirring. During the addition, silver 5-nitrotetrazolate gradually precipitated from the reaction mixture as a white solid. After the addition was complete, the suspension was stirred for an additional 30 min. The product was subsequently collected by filtration and washed thoroughly with water. After air-drying overnight, silver 5-nitrotetrazolate was obtained as a colorless powder.~\cite{Lommel_2025}

Analysis: FT-IR (cm$^{-1}$): 1632w, 1548w, 1455w, 1396s, 1304w, 1185m, 1169m, 1085w, 1048m, 1032w, 878w, 831w. Elemental analysis: Calc. for CN$_5$O$_2$Ag (\%): C 5.41, N 31.56. Found: C 5.24, N 31.21. DTA (5~$^\circ$C\,min$^{-1}$): 230$^\circ$C (start of decomposition), 273$^\circ$C (maximum). 

\subsubsection{Silver 5,5'-azotetrazolate}
Sodium 5,5‘-azotetrazolate pentahydrate (150 mg, 0.5 mmol) was dissolved in 5 mL of water in a plastic beaker wrapped in aluminum foil to protect the reaction mixture from light. A solution of silver nitrate (204 mg, 1.2 mmol) in 5 mL of water was then added dropwise under continuous stirring. During the addition, silver 5,5' -azotetrazolate gradually precipitated from the reaction mixture as a orange solid. After the addition was complete, the suspension was stirred for an additional 15 min. The product was subsequently collected by filtration and washed twice with 5 mL water and twice with 5 mL ethanol. After air-drying overnight, silver 5,5‘-azotetrazolate was obtained as a dark-orange powder. Analysis: FT-IR (cm$^{-1}$): 3382b, 1622w, 1538w, 1463vw, 1452vw, 1436vw, 1397s, 1315m, 1211m, 1185m, 1170w, 1072vw, 1047m, 1033m, 921w, 837w, 792vw, 766m, 735s, 719s. Elemental analysis: Calc. for C$_2$N$_{10}$Ag$_2$ (\%): C 6.32, N 36.88. Found: C 6.55, H 0.30, N 36.16. DTA (5~$^\circ$C\,min$^{-1}$): 230°C (start of decomposition), 273°C (maximum). 
\subsubsection{Silver fulminate}
The procedure for the preparation of silver fulminate was based on those reported in the literature by Taylor \textit{et al.}~\cite{Taylor1926} and Fedoroff \textit{et al.}~\cite{Fedoroff1974} Silver nitrate (1.70 g, 10 mmol) was dissolved in 65\% nitric acid (9 ml, 130 mmol) and 10 ml of water. After dissolving the silver nitrate, this solution was added to 18.5 ml (300~mmol) of 95\% ethanol and the mixture was heated to 80$^{\circ}$C, at which point a white precipitate of silver fulminate began to form. After 30 minutes, the silver fulminate was filtered, washed several times with water, and allowed to dry at room temperature for 24 h.

\subsubsection{Silver acetylide}
Silver acetylide was prepared according to the Stettbacher procedure.~\cite{Stettbacher1916Neuere} Silver nitrate (1.70 g, 10 mmol) was dissolved in 14 ml water; then 15 ml of 25\% aqueous ammonia (200 mmol) was added to the silver nitrate solution. After dissolving the temporarily formed white precipitate, with a further addition of ammonia, pure acetylene from a gas cylinder was introduced into the solution. The silver acetylide immediately formed as a white precipitate. This was then isolated by filtration, washed thoroughly with water and allowed to dry for 24 h at room temperature.~\cite{Klap_tke_2009}

Analysis: FT-IR (cm$^{-1}$): 3345m, 2920m, 2851m, 2173m, 1372s, 1030m, 830m, 666m. Elemental analysis: Calc. for C$_2$Ag$_2$ (\%): C 10.02, Ag 89.98. Found: C 10.34, H 0.23, N 0.12. DTA (5$^{\circ}$C min$^{-1}$): 122$^{\circ}$C (start of decomposition), 160$^{\circ}$C (maximum).

\subsubsection{Cadmium 5,5'-azotetrazolate complex}
Cadmium 5,5'-azotetrazolate complex was prepared by precipitation from aqueous solutions of sodium 5,5'-azotetrazolate pentahydrate (300 mg, 1 mmol) in water (10 ml) and cadmium nitrate tetrahydrate (308 mg, 1 mmol) in water (5 ml). Upon mixing the solutions under continuous stirring, an immediate yellow precipitate formed. The resulting suspension was stirred for an additional 10 min and then filtered. The solid product was washed thoroughly with distilled water and allowed to dry. Cadmium 5,5'-azotetrazolate complex was obtained as a yellow powder.~\cite{Qiu_2013}

Analysis: FT-IR (cm$^{-1}$): 3396br, 2478w, 2278w, 2143w, 1621m, 1499vw, 1475vw, 1456vw, 1406s, 1217m, 1189m, 1171w, 1103w, 1073w, 1039m, 770w, 749s.
Elemental analysis: Calc. For C$_4$H$_{10}$Cd$_2$N$_{20}$O$_5$ (\%): C 7.47, H 1.57, N 43.56, O 12.36. Found: C 7.87, H 1.80, N 43.69. 
DTA (5~$^\circ$C\,min$^{-1}$): 143~$^\circ$C (start of decomposition), 169~$^\circ$C (maximum).
Metal analysis: calculated Cd 34.96\%; found Cd 36.10\%.
\subsubsection{Palladium 5,5'-azotetrazolate}
Palladium 5,5'-azotetrazolate was prepared by precipitation from aqueous solutions of sodium 5,5'-azotetrazolate pentahydrate (300 mg, 1 mmol) in water (10 ml) and palladium nitrate  (300 mg) in water (5 ml). Upon mixing the solutions under continuous stirring, an immediate yellow precipitate formed. The resulting suspension was stirred for an additional 10 min and then filtered. The solid product was washed thoroughly with distilled water and allowed to dry. Palladium 5,5'-azotetrazolate was obtained as a yellow powder.

\subsection{Detonation and particle collection}
BANG experiments were performed using a purpose-built stainless-steel collection device designed to direct the products of detonation towards a collection substrate. The same collection device was used for the Pb, Bi, Cd, Tl, Mn, Fe, Ni, Cu, Zn, Pt and Pd experiments.

Typically, between 10 and 20~mg of energetic material was used for each detonation experiment. For experiments performed under ambient atmosphere, detonation was initiated by heating the energetic material using a gas torch. A holey-carbon film on a 400-mesh copper/nickel TEM grid was positioned above the energetic material for particle collection. The grid was positioned approximately 10--20~cm above the detonation region. 

\subsection{Nitrogen-atmosphere BANG experiment}\label{sec:nitrogenBANG}
To investigate nanoparticle formation under an inert atmosphere, silver azide was detonated under nitrogen. The stainless-steel collection device, piezoelectric igniter and laboratory-prepared silver azide were placed inside a glove bag, which was filled with nitrogen gas before initiation. Between 10 and 20~mg of silver azide was used. In contrast to experiments performed under ambient atmosphere, initiation was achieved using a piezoelectric spark rather than heating with a gas torch. Particles generated during detonation were collected on an electron-microscopy grid positioned approximately 10--20~cm above the energetic material.

\subsection{Collection on macroscopic substrates}\label{sec:depo}
The same stainless-steel cylinder was also used for experiments in which material was deposited directly onto macroscopic substrates. Typically, between 10 and 20~mg of energetic precursor was used, and experiments under ambient atmosphere were initiated using a gas torch. For palladium deposition experiments, activated carbon was positioned in the path of material generated during detonation of palladium 5,5'-azotetrazolate, resulting in direct deposition of Pd-containing nanoparticles onto the carbon support.

\subsection{DTA}
Differential thermal analysis was performed using a DTA 550Ex differential thermal analyser (OZM Research, Czech Republic) with calibration on indium. The samples were tested in open glass microtest tubes in contact with air at a heating rate of 5 $^\circ$C min$^{-1}$. The sample weights were 5 mg.
\subsection{Metal Content Analysis}
The metal contents of selected energetic precursors were determined by flame atomic absorption spectrometry (FAAS) using an Avanta AA 906 atomic absorption spectrometer (GBC) equipped with element-specific hollow-cathode lamps and operated with an air--acetylene flame. The instrumental parameters were adjusted for each metal. Samples were dissolved in dilute nitric acid prepared by mixing 10 ml of nitric acid with 100 ml of water. Each sample was analysed in triplicate, and the metal content was determined from the corresponding absorbance--concentration calibration curve.
\subsection{CHNS Elemental Analysis}
Elemental analysis was carried out using a UNICUBE automatic elemental analyser (Elementar, Germany) in CHNS mode with calibration on sulphanilamide, using 0.5-1.5~mg samples for this analysis 
\subsection{IR}
Infrared spectra were collected using a Nicolet iS50 FT-IR spectrometer (Thermo Fisher, USA) with an iS50ATR diamond and ZnSe accessory Smart iTr (Thermo Fisher Scientific, USA). Measurement parameters were the spectral regions of 4000-400 cm$^{-1}$ for iS50ATR diamond and 4000-600 cm$^{-1}$ for ZnSe; the resolution of 4 cm$^{-1}$, and 64 scans.
\subsection{Electron Microscopy}
Particles were analysed using a JEOL JEM-2100 TEM and JEOL JSM-IT300 SEM. Electron diffraction analysis was performed with CrysTBox.~\cite{Box2,Box1,Box3}
The Inorganic Crystal Structure Database (ICSD) collection codes for the reference structures used in the crystallographic analysis are given in~\autoref{tab:collectionCodes}.
\begin{table}[htbp]
    \centering
    \caption{Inorganic Crystal Structure Database (ICSD) collection codes for reference structures used in the crystallographic analysis.}
    \label{tab:collectionCodes}
    \begin{tabular}{c|c}
        Phase & ICSD Collection Code \\
        \hline
        Pb$_3$O$_5$ & 36294~\cite{Bystrom1945Decomposition}\\
        Pb$_3$O$_4$ & 4106~\cite{Gavarri_1975}\\
        Bi$_2$O$_3$ & 15072~\cite{Malmros_1970}\\
        (Tl$_2$O$_3$)$_{0.89}$ & 54214~\cite{Lyons_2002}\\
        Tl$_4$O$_3$ & 23478~\cite{Marchand1973Tl4O3}\\
        Mn$_3$O$_4$ & 33327~\cite{Rajeesh_Kumar_2021}\\
        Fe$_4$O$_5$ & 47905~\cite{Layek_2022}\\
        CuO & 16025~\cite{Aasbrink1970}\\
        Cu$_4$O$_3$ & 77675~\cite{Morgan_1996}\\
        ZnO$_2$ & 38979~\cite{Bocharov_2022}\\
        PtO$_2$ & 251568~\cite{Shirako_2014}\\
        NiO & 9866~\cite{SASAKI_1979}\\
        Ag$_3$O$_4$ & 202218~\cite{Standke_1987}\\
        AgO & 202543~\cite{Jansen_1988}\\
        HCP Ag & 56269~\cite{Novgorodova1979,Novgorodova1981}\\
        Cd & 619639~\cite{Rossmanith_1978}\\
    \end{tabular}
\end{table}
\section{Safety Statement}
The metal-containing energetic precursors investigated in this work include primary explosives and other highly sensitive energetic materials. All preparation, handling and detonation experiments were therefore performed on small scales under approved institutional risk assessments and standard operating procedures. The quantity of energetic material used in individual BANG experiments was typically limited to 10--20~mg.

Energetic materials were handled within a fume hood using appropriate engineering controls and personal protective equipment. Electrostatic-discharge precautions were employed where appropriate, and unnecessary confinement or manual restraint of energetic materials was avoided. Detonation experiments were performed using a purpose-built stainless-steel collection device intended to provide controlled directional collection of the detonation products.

Heavy-metal-containing, azide-containing and other hazardous waste streams were treated and disposed of according to approved institutional procedures.

\textbf{Caution:} The energetic compounds described in this work can be sensitive to heat, impact, friction and electrostatic discharge. Their preparation and handling should only be undertaken by appropriately trained personnel using suitable facilities, local risk assessments and procedures approved for energetic materials.

\section{Conclusions}
The work presented here significantly expands BANG beyond the initially reported Au, Ag and Cu systems and demonstrates its broader chemical scope. Detonation of metal-containing energetic precursors results in the formation of highly crystalline, metal-containing nanoparticles. The resulting products depend strongly on the chemistry of the metal and the detonation environment. More oxophilic metals predominantly formed oxide-containing nanoparticles following detonation in air, whereas detonation of silver azide under nitrogen suppressed detectable silver oxide formation.  Local HRTEM and FFT analysis additionally identified regions consistent with metastable HCP Ag, supporting the ability of the rapid, highly non-equilibrium BANG environment to retain structures that are uncommon under equilibrium conditions. This mirrors behaviour reported for EEW, where noble metals exploded in air can form metallic nanoparticles, whereas base metals commonly yield metal/oxide mixtures and require inert atmospheres to preserve metallic products.~\cite{Kotov_2003} This similarity to EEW opens up a future possibility of expanding the synthesis to bimetallic or multimetallic particles.

A more focused study of the nickel systems showed that variation in the energetic precursor had no noticeable effect on the resulting phase, with all three precursors producing NiO. The palladium system further demonstrated that BANG-derived nanoparticles can be deposited directly onto a substrate during synthesis. The characteristic features of BANG-derived particles include broad particle-size distributions, aggregation and limited morphological control, reflecting the rapid, highly non-equilibrium nature of plume-based particle formation. Nevertheless, these results establish BANG as a broadly applicable, chemically driven high-energy synthesis route and provide a basis for future control over composition, phase and deposition through precursor and atmosphere selection. 

\subsection*{Author Contributions}
\textbf{Conceptualization:} JMU, SRH. \\
\textbf{Methodology:} JMU.\\
\textbf{Investigation:} JMU,RM.\\
\textbf{Visualization:} JMU, SRH.\\
\textbf{Funding acquisition:} SJE, NS.\\
\textbf{Project administration:} JMU, SRH, SJE, AJP, NS.\\
\textbf{Supervision:} SRH, SJE, AJP, NS.\\
\textbf{Writing -- original draft:} JMU, RM, SRH.\\
\textbf{Writing -- review and editing:} JMU, SRH, SJE, AJP, RM.


\subsection*{Conflicts of Interest}

The authors declare no conflicts of interest.

\subsection*{Data Availability Statement}
Data supplied by the corresponding author upon request.



\end{document}